\documentclass[aps,prx,twocolumn,colorlinks=true,linkcolor=black,urlcolor=blue,citecolor=black,pdfencoding=auto,longbibliography,superscriptaddress]{revtex4-2}
\usepackage{graphicx}
\usepackage{color}
\usepackage{amsmath}
\usepackage{amsfonts}
\usepackage{bbold}
\usepackage{amssymb}
\usepackage{tikz}
\usepackage{bm}
\usepackage{braket}
\usepackage{pgfplots}
\usepackage{dsfont}
\usepackage{comment}
\usepackage[colorlinks=true,
            citecolor=blue,
            pdfencoding=auto]{hyperref}
\usepackage{blindtext}
\usepackage{mathrsfs}
\usepackage[scr=boondoxo,scrscaled=1.05]{mathalfa}

\makeatletter
\newcommand{\overleftrightsmallarrow}{\mathpalette{\overarrowsmall@\leftrightarrowfill@}}
\newcommand{\overrightsmallarrow}{\mathpalette{\overarrowsmall@\rightarrowfill@}}
\newcommand{\overleftsmallarrow}{\mathpalette{\overarrowsmall@\leftarrowfill@}}
\newcommand{\overarrowsmall@}[3]{%
  \vbox{%
    \ialign{%
      ##\crcr
      #1{\smaller@style{#2}}\crcr
      \noalign{\nointerlineskip}%
      $\m@th\hfil#2#3\hfil$\crcr
    }%
  }%
}
\def\smaller@style#1{%
  \ifx#1\displaystyle\scriptstyle\else
    \ifx#1\textstyle\scriptstyle\else
      \scriptscriptstyle
    \fi
  \fi
}
\makeatother
\newcommand{\te}[1]{\overleftrightsmallarrow{#1}}

\begin{document}

%\preprint{APS/123-QED}
\title{Theory of Topological Nernst and Thermoelectric Transport in Chiral Magnets}

\author{Zachariah Addison}%
    \thanks{These authors contributed equally}
\affiliation{Department of Physics,
Ohio State University,
Columbus, OH 43210, USA}
 \author{Lauren Keyes}%
\thanks{These authors contributed equally}
\affiliation{Department of Physics,
Ohio State University,
Columbus, OH 43210, USA}
\author{Mohit Randeria}
\affiliation{Department of Physics,
Ohio State University,
Columbus, OH 43210, USA}

\date{\today}% It is always \today, today,

\begin{abstract}
We calculate the thermoelectric transport of spin-orbit coupled conduction electrons in the presence of topological spin textures. We show, within a controlled, semiclassical approach that includes all phase space Berry curvatures, that the Nernst effect has two contributions in addition to the usual effect proportional to a magnetic field. These are an anomalous contribution governed by the momentum-space Berry curvature and proportional to net magnetization, and a topological contribution determined by the real-space Berry curvature and proportional to the topological charge density, which is non-zero in skyrmion phases. We derive a generalized Mott relation expressing the thermoelectric tensor as the chemical potential derivative of the conductivity tensor and show how the Sondheimer cancellation in the Nernst effect is evaded in chiral magnets.
\end{abstract}

\maketitle
There has been enormous effort  in the investigation of chiral magnetic materials 
in recent years \cite{roessler2006spontaneous, neubauer2009topological, nagaosa2013topological, fert2017magnetic}.
This has been in part due to the fundamental interest in topological spin textures and their impact on the 
properties of materials, and in part motivated by the possibility of using skyrmions (topological textures of unit charge) for potential device applications.

One of the most widely studied effects of topological charge density in chiral magnets is their unusual signature in transport: the topological Hall effect (THE)~\cite{nagaosa2013topological}. This effect arises when the conduction electrons -- in metallic magnets~\cite{lee2009unusual, neubauer2009topological, kanazawa2011large, li2013robust, gallagher2017robust, ahmed2018chiral} or in heavy metals proximate to a magnet~\cite{ahmed2019spin, shao2019topological} --
are impacted by an ``emergent magnetic field", which is the flux quantum $(h/e)$ times the topological charge, 
density \cite{ye1999berry, bruno2004topological, nagaosa2012gauge, nagaosa2013topological, kim2013chirality, akosa2019tuning}. 
Our focus here is the analogous topological effect in the transverse {\it thermoelectric} response. 
%MR: Do we really need all these THE references?

The Nernst signal $N$, the transverse voltage response to an applied thermal gradient in the absence time reversal symmetry, is a 
quantity of fundamental importance. It is well known that $N = - E_y/|\nabla_x T| $ in an external magnetic field $B_z$ is vanishingly small in
simple metals due to the Sondheimer cancellation \cite{sondheimer1948theory, wang2001onset}. A large Nernst effect is usually observed in either semimetals or
strongly correlated systems \cite{behnia2016nernst}.

Naively, if the ``emergent magnetic field" due to a nontrivial topological charge density was simply analogous to an external 
magnetic field (as in the theory of the THE~\cite{bruno2004topological}) one might expect a Sondheimer cancellation
and a very small topological Nernst effect.
It is thus interesting that a robust topological Nernst effect has been seen in the 
skyrmion phase of chiral magnets~\cite{shiomi2013topological,   hirschberger2020topological, kolincio2021large, 
scarioni2021thermoelectric, macy2021magnetic, zhang2021topological}.

In this paper we develop a theory of the topological Nernst effect in chiral magnetic materials that addresses this puzzle. In addition, we also need to address the issue that the topological contribution is only one part of the observed signal. 
 Experiments~\cite{shiomi2013topological,   hirschberger2020topological, kolincio2021large, 
scarioni2021thermoelectric, macy2021magnetic, zhang2021topological}
on the transverse thermoelectric response in chiral magnetic are
analyzed as the sum of three pieces: an ``ordinary" response proportional to the external magnetic field, an ``anomalous" contribution proportional to the magnetization,
and a ``topological" contribution proportional to the topological charge density $n_{\text{top}}=\int d^3r \, \hat{\bm{m}}\!\cdot\!(\partial_{r_x}\hat{\bm{m}}\!\times\!\partial_{r_y}\hat{\bm{m}})/4\pi V$. This decomposition is motivated by the empirical success of a similar expression for Hall resistivity~\cite{lee2009unusual, neubauer2009topological, kanazawa2011large, li2013robust, gallagher2017robust, ahmed2018chiral,
ahmed2019spin, shao2019topological} as the sum of three contributions.

The results presented here build on the recent demonstration~\cite{verma2022unified} that, within a controlled semiclassical calculation,
the Hall response arising from chiral magnetism can be shown rigorously to be the sum of an (intrinsic) anomalous contribution, proportional to the 
${\bf k}$-space Berry curvature \cite{nagaosa2010anomalous,xiao2010berry} and a topological contribution, proportional to the 
real-space Berry curvature. We analyze the dynamics of wave-packets in phase space, taking
into account {\it all} Berry curvatures (including the mixed $({\bf r},{\bf k})$ curvatures) on an equal footing
together with ${\bf r}$ and ${\bf k}$ derivatives of the semiclassical energy eigenvalues.
In the semiclassical regime where the lattice spacing $a \ll$ the mean free path $\ell \ll L_s$ the spin texture length scale, and weak
SOC $\lambda$ compared to electronic energy scales, we solve the Boltzmann equation to 
determine the thermoelectric tensor $\te{\alpha}$, which relates the electrical transport current $\bm{j}_{\text{tr}}$ to 
the temperature gradient via $\bm{j}_{\text{tr}}=-\te{\alpha}\bm{\nabla}_r T$. 

 \begin{figure*}
\includegraphics[width=\textwidth]{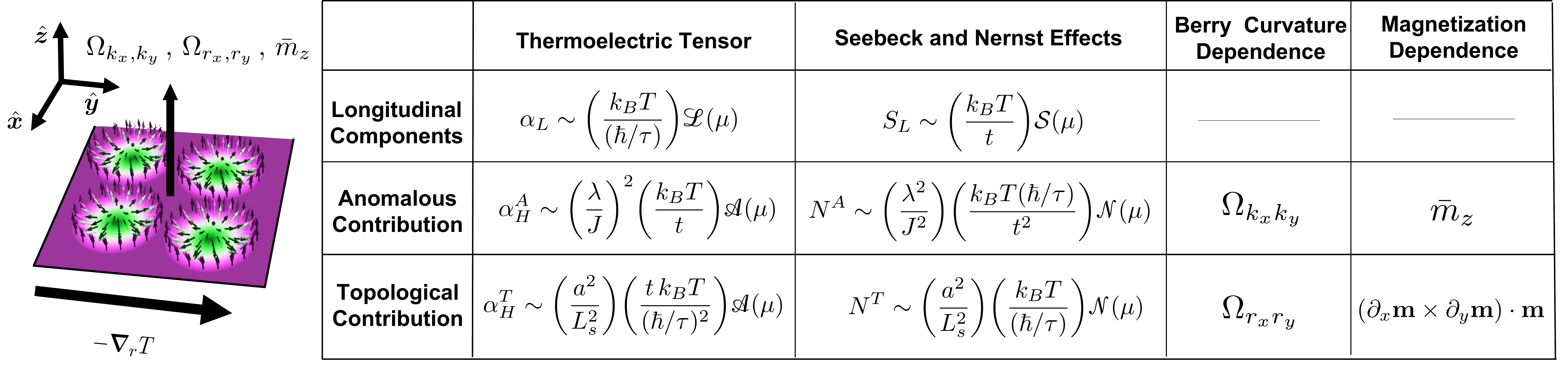}
\caption{{\bf Summary of Results.} Dominant scaling relations, Berry curvature and magnetization dependencies for leading order contributions to the thermoelectric conductivity, Seebeck, and Nernst effects in the regime $k_F^{-1}\sim a\ll\ell=v_F\tau\ll L_s$ and $\lambda \ll J < t\sim E_F$.  The Seebeck and Nernst effects are related to the thermopower tensor $\protect\te{S}=\protect\te{\sigma}^{-1}\protect\te{\alpha}$, where Seebeck $S_L=1/3(S_{xx}+S_{yy}+S_{zz})$ and Nernst $N=(S_{xy}-S_{yx})/2$. The chemical potential or density dependence of quantities are determined by the dimensionless functions $\mathscr{L}(\mu)$, $\mathcal{S}(\mu)$, $\mathscr{A}(\mu)$, and $\mathscr{N}(\mu)$ that depend on the ratios $J/\mu$ and $J/t$.}
 \label{tab-scaling}
\end{figure*}

We summarize our main results.
\\
(1) We show that, to leading order in the small parameters indicated above, the transverse (off-diagonal) thermoelectric response 
in a system with spin textures is just the sum of an anomalous piece and a topological piece. As summarized in the Table in Fig.~\ref{tab-scaling},
the former arises from ${\bf k}$-space Berry curvature~\cite{xiao2006berry} and is proportional to the net magnetization, 
while the latter arises from ${\bf r}$-space Berry curvature and is proportional to the topological charge density. All other contributions,
arising, e.g., from mixed curvatures, are small corrections in the semiclassical regime with weak SOC.
\\
(2) We derive a Mott relation relating the thermoelectric tensor $\te{\alpha}$ to the chemical potential derivative $\partial \te{\sigma}(\mu)/\partial \mu$
of the electric conductivity tensor $\te{\sigma}$. A Mott relation for just the anomalous response in a ferromagnet was derived in the pioneering work of Ref.~\cite{xiao2006berry}; here we show that it is valid in the presence of arbitrary spin textures including both the anomalous and topological terms.
\\
(3)
We show how the topological Nernst contribution evades the Sondheimer cancellation. The ${\bf r}$-space Berry curvature couples with opposite signs to the spin-split conduction bands, unlike an external magnetic field, and this leads to a non-zero contribution even for a simple parabolic dispersion.
\\
(4) Although the anomalous and topological contributions originate from vastly different physical mechanisms, we find, somewhat surprisingly, that they
have the same functional dependence on the chemical potential $\mu$ or density of conduction electrons, provided the SOC
is proportional to the conduction electron group velocity.

Our conclusions are derived for conduction electrons with arbitrary dispersion and a general form for the SOC, including
Rashba SOC arising at interfaces, interacting with any spin texture in 2D. More generally, we also analyze the 3D problem 
with a spin texture that does not vary in the $z$-direction, as would be the case for a random array or a crystal of skyrmion tubes.

Previous theoretical analyses of thermoelectric transport 
%MR
in chiral magnets 
have been restricted to either numerical calculations~\cite{mizuta2016large}, where a 
decomposition into the anomalous and topological contributions is ill-defined, or to an analytic approach \cite{ gayles2018topological}
that ignores SOC so that spin remains a good quantum number and, in addition, the Mott relation is assumed rather than derived.

%MR: Cite later? mixed phase space Berry curvatures \cite{sundaram1999wave, xiao2010berry, verma2022unified}.  

\medskip
\noindent
%MR: section rewritten -- also removed H_{KE} etc. in (1) because it is not correct to include J term in KE
\textbf{Model:} We consider the Hamiltonian
\begin{align}
\widehat{H} &= -\sum_{ij,\sigma}t_{ij}\hat{c}^{\dagger}_{i\sigma}\hat{c}^{\phantom{\dagger}}_{j\sigma} -J \sum_{i,\sigma\sigma'}\hat{c}^{\dagger}_{i\sigma}(\hat{\bm{m}}(\bm{r}_i)\cdot \boldsymbol{\sigma}^{\sigma\sigma'})\hat{c}^{\phantom{\dagger}}_{i\sigma'} \nonumber\\
&+ \dfrac{\lambda\hbar}{at} \sum_{ ij,\gamma\delta,\sigma\sigma'}\hat{c}^{\dagger}_{i\sigma}
(v_\gamma^{ij}\,\chi^{\phantom{\dagger}}_{\gamma\delta}\sigma_\delta^{\sigma\sigma'})\hat{c}^{\phantom{\dagger}}_{j\sigma'}
\label{eq:ham-gen}
\end{align}
\noindent
where $i,j$ label lattice sites, $\sigma,\sigma'\in \{\uparrow,\downarrow\}$  and $\gamma,\delta\in \{x,y\}$.
The first term describes an arbitrary band structure using tight-binding amplitudes
$t_{ij}$ whose scale is $t$. The second term couples the conduction electron spin to a given
magnetic texture $\hat{\bm{m}}(\bm{r})$ with an exchange coupling $J$. For simplicity we choose
$\hat{\bm{m}}(\bm{r})$ to be independent of $z$, which is adequate to model crystals or disordered arrays of
skyrmion tubes.

The SOC with strength $\lambda$ is proportional to the electron velocity $\bm{v}^{ij}=it_{ij}(\bm{r}_i-\bm{r}_j)/\hbar$ 
on a bond with lattice constant $a$.  For simplicity, we restrict ourselves to SOC that involves
only $\sigma_x$ and $\sigma_y$ as appropriate for systems with broken interfacial inversion.
The precise form of the SOC depends on the $\te{\chi}$ tensor. $\te{\chi}=i\tau_y$ leads to Rashba SOC 
$({v}^{ij}_x\sigma_y - {v}^{ij}_y\sigma_x)$ which preserves vertical mirror planes 
($\mathcal{M}_x$, $\mathcal{M}_y$), but breaks $\mathcal{M}_z$.  
Choosing $\te{\chi}=\tau_z$ leads to $({v}^{ij}_x\sigma_x - {v}^{ij}_y\sigma_y)$ which breaks all mirror 
planes~\cite{dresselhaus1955spin,bychkov1984properties}. 
(The effects of Ising SOC $\propto\sigma_z$ are suppressed by $\lambda/J$ and ignored; see Appendix \ref{appA}). 

Finally, we include effects due to impurity scattering processes in $\widehat{H}_{\text{imp}}$ which will enter our Boltzmann equation 
analysis below through the relaxation time $\tau$. The energy scales in our model can be organized as
$\lambda \ll J < t \sim E_F$, where $E_F$ is the Fermi energy measured from the band edge and where
$E_F\gg k_B T$.

To take into account both ${\bf r}$ and ${\bf k}$-space Berry curvatures at the same time, we need to use a semiclassical approach. 
This demands that the microscopic length scales $a \sim k_F^{-1}$ are much smaller than the mean free path $\ell = v_F \tau$
and the length scale $L_s$ on which the spin texture varies. To control our calculations we will work in the regime 
 $a \ll \ell\ll L_s$. These are realistic assumptions for many chiral magnetic materials, where 
 $10 \lesssim L_s \lesssim 500$ nm~\cite{tokura2020magnetic}, while $1 \lesssim \ell \lesssim 100$ nm
 (given that $10 \lesssim k_F\ell \lesssim 100$).

\medskip
\noindent
\textbf{Semiclassical Theory of Thermoelectric Transport:} To analyze the dynamics of wave packets in phase space
$\bm{\xi}=(r_x,r_y,r_z,k_x,k_y,k_z)$, we construct the semiclassical Bloch Hamiltonian 
$\mathcal{H}_{sc}(\boldsymbol\xi) = \varepsilon(\bm{k})\mathbb{1}+ {\bf d}(\boldsymbol\xi) \cdot {\boldsymbol\sigma}$, where
\begin{align}
 d_\gamma(\boldsymbol\xi) &= \dfrac{\lambda}{a\,t}\,\sum_{\delta}\chi^{\phantom{\dagger}}_{\gamma\delta}\partial_{k_\delta}\varepsilon(\bm{k}) - J \hat{m}_\gamma({\bf r}); \,\,\, \gamma,\delta\in \{x,y\} \nonumber\\ d_z(\bm{\xi})&=-J\hat{m}_z(\bm{r})
\label{ham-sc-1}
\end{align}
\noindent
where $\varepsilon(\bm{k})$ is the band dispersion in the absence of $\lambda$ and ${\bf d}(\boldsymbol\xi)$ captures the 
quantum mechanical nature of the spin (see Appendix \ref{appA} for details). The semiclassical eigenenergies are 
$\mathcal{E}_\pm(\boldsymbol\xi) =\varepsilon(\bm{k}) \pm | {\bf d} (\boldsymbol\xi) |$ and the derivatives of the eigenfunctions, $\ket{u_l(\bm{\xi})}$, encode the quantum geometry of the semiclassical bands through the generalized Berry curvatures
\begin{eqnarray}
\Omega^\pm_{ \alpha\beta }(\boldsymbol\xi) &=& \pm \dfrac{1}{2} \hat{ {\bf d} }(\boldsymbol\xi) \cdot \left( \partial_{\alpha } \hat{\bf d}(\boldsymbol\xi) \times \partial_{\beta } \hat{{\bf d}}(\boldsymbol\xi) \right)
\end{eqnarray}
\noindent
with $\alpha,\beta \in\{r_x,r_y,r_z,k_x,k_y,k_z\}$.  
The semiclassical equations of motion (with band index $l= \pm$) are
\begin{equation}
\dot{\xi}_\alpha^l (\boldsymbol\xi) = [\Gamma^{-1}_l(\boldsymbol\xi)]_{\alpha\beta} \; \partial_{\beta} \widetilde{\mathcal{E}}_l(\boldsymbol\xi)/\hbar \label{eq:eom-Gamma}
\end{equation}
\noindent
where $[\Gamma_l(\boldsymbol\xi)]_{\alpha\beta} = \Omega^l_{ \alpha \beta}(\boldsymbol\xi)-[ i\sigma_y \otimes \mathds{1} ]_{\alpha\beta}$.
$\widetilde{\mathcal{E}}_l(\boldsymbol\xi) \simeq \mathcal{E}_l(\boldsymbol\xi)$ up to corrections of order $(\lambda/E_F)(a/L_s)$ that can be ignored in the regime of interest  \cite{xiao2005berry,verma2022unified}.  We  suppress the $\bm{\xi}$-dependence of quantities in what follows.

Building on the analysis of Ref. \cite{xiao2010berry} we obtain the local charge current
\begin{align}
\bm{j}_{\text{loc}}&=\sum_{l=\pm}\int \dfrac{d^3k}{(2\pi)^3} \bigg(-e\mathcal{D}_lf_l\dot{\bm{r}}_l+\bm{\nabla}_r\times(\mathcal{D}_lf_l\bm{\mathfrak{m}}_l)\bigg)
\label{curr}
\end{align}
\noindent
The first term describes the center of mass motion of wave packets, while the second describes their orbital rotation.  
Here $\dot{\bm{r}}_l$ is determined by Eq.~\eqref{eq:eom-Gamma}, $f_l$ is the electronic distribution function, and
$\mathcal{D}_l =\sqrt{\det[\Gamma_l(\boldsymbol \xi )]}$ describes the modification of the phase space volume element in the presence of Berry curvatures so that Liouville's theorem is satisfied. The orbital magnetic moment $\bm{\mathfrak{m}}_l$ of the semiclassical wave packet (with $a,b,c\!\in\!\{x,y,z\}$) is given by
\begin{align}
\bm{\mathfrak{m}}_l\!\cdot\!\hat{\bm{r}}_a&=-i\dfrac{e}{2\hbar}\sum_{bc}\varepsilon_{abc} (\partial_{k_b}\bra{u_l})(\mathcal{H}_{sc}-\mathcal{E}_l)(\partial_{k_c}\ket{u_l})
\end{align}
\noindent

In experiments, the Nernst effect is determined by the $\bm{q}=0$ {\it transport} current~\cite{cooper1997thermoelectric}
\begin{equation}
\bm{j}_\text{tr}=\int \dfrac{d^3r}{V}\bigg( \bm{j}_{\text{loc}}-\bm{\nabla}_r\times\bm{M}\bigg)
\label{currTr}
\end{equation}
\noindent
where $V$ is the volume of the system and $\bm{M}$ the thermodynamic magnetization.  To calculate anomalous and topological contributions to the thermoelectric conductivity, we expand Eq.~\eqref{currTr} to first order in temperature gradients (linear response) and to second order in the small parameters of our theory. 

\medskip
\noindent
\textbf{Thermoelectric Conductivity.}  We write the thermoelectric tensor as $\te{\alpha}=\te{\alpha}^{(1)}+\te{\alpha}^{(2)}$, 
where $\alpha^{(1)}$ is independent of $\tau$, and $\alpha^{(2)}$ depends on $\tau$.  
$\te{\alpha}^{(1)}$ arises from the orbital magnetic moment in E.~\eqref{curr} and the magnetization in Eq.~\eqref{currTr}.  
The antisymmetric Hall component $\alpha_H=(\alpha_{xy}-\alpha_{yx})/2$ has two leading order contributions, $\alpha_H^A$ and $\alpha_H^T$. Here, $\alpha^A_H=(\alpha^{(1)}_{xy}-\alpha^{(1)}_{yx})/2$ is the anomalous contribution to the thermoelectric 
Hall conductivity~\cite{xiao2006berry}, given by
\begin{equation}
    \alpha^A_{H}=\dfrac{\lambda^2}{T} \int_\xi ((\varepsilon_l-\mu)f^0[\varepsilon_l]-G[\varepsilon_l])\bigg(\dfrac{\partial^2 \Omega^l_{k_xk_y}}{\partial \lambda^2}\bigg)\bigg|_{\lambda=0}
    \label{TEA}
\end{equation}
\noindent
Here $\int_\xi\equiv \sum_{l=\pm}\int d^6\xi/(8\pi^3 V)$, the local grand potential density
$G_l[\varepsilon_l]=-k_BT\ln(1+e^{-\beta(\varepsilon_l-\mu)})$, 
the equilibrium distribution function $f^0[\varepsilon_l]=(e^{(\varepsilon_l-\mu)/k_B T}+1)^{-1}$, 
and $\varepsilon_l$ are the semiclassical eigenenergies in the absence of $\lambda$: $\varepsilon_\pm(\bm{k})=\varepsilon(\bm{k})\pm J$.  
We note that real space gradient corrections to Eq.~\eqref{TEA} are down by $a/L_s$.

The $\tau$-dependent contribution $\te{\alpha}^{(2)}$ is obtained from the Boltzmann equation
\begin{equation}
    -\dfrac{f_l-f^0[\mathcal{E}_l]}{\tau}=\dot{\bm{r}}\cdot\bm{\nabla}_r f_l+\dot{\bm{k}}\cdot\bm{\nabla}_k f_l
    \label{boltz}
\end{equation}
which we solve for $f_l =  f^0[\mathcal{E}_l]+g_l$ to linear order in the temperature gradient 
witin the relaxation time approximation (see Appendix \ref{appB}).  
We find that the leading order longitudinal contribution $\alpha_{L}=(\alpha_{xx}^{(2)}+\alpha_{yy}^{(2)}+\alpha_{zz}^{(2)})/3$ 
can be written as
\begin{equation}
\alpha_{L}=-\dfrac{e\tau}{3\hbar^2} \int_\xi\,|\bm{\nabla}_k\varepsilon_l|^2\,\partial_Tf^0[\varepsilon_l]
\label{alphaL}
\end{equation}

\noindent
The topological Hall contribution to the theormoelectric conductivity derives from the antisymmetric component 
$\alpha_H^T=(\alpha^{(2)}_{xy}-\alpha^{(2)}_{yx})/2$ that can be written as
\begin{align}
    \alpha_H^{T}&=
\dfrac{e\tau^2 t^3}{\hbar^3}\,n_{\text{top}}a^2 \sum_{l=\pm}\, l\,\int \dfrac{d^3k}{8\pi^2}\,\partial_{T}f^0[\varepsilon_l]\,\mathscr{Q}(\bm{k}).
\label{trueAlpha}
\end{align}
\noindent
Here $n_{\text{top}}$ is the topological charge density and 
\begin{equation}
\mathscr{Q}(\bm{k})= \dfrac{1}{(a^2\, t^3)} \bigg(\bm{v}^T\cdot(\te{M}^{-1}-{\rm Tr}(\te{M}^{-1})\mathds{1})\cdot\bm{v}\bigg)
\label{Q}
    \end{equation}
\noindent
with $\bm{v}=(\partial_{k_x}\varepsilon,\partial_{k_y}\varepsilon$) and $\te{M}^{-1}_{k_ik_j}=\partial_{k_i}\partial_{k_j} \varepsilon_l$ (see Appendix \ref{appC}).   

Thus, the leading order contributions to the thermoelectric conductivity are just the sum of an anomalous contribution proportional to the momentum space Berry curvature and a topological contribution proportional to the topological charge density: $\alpha_H = \alpha^A_H + \alpha^T_H$ with corrections suppressed in powers of the small parameters of our theory  (see Fig. \ref{tab-scaling}).

\medskip
\noindent
\textbf{Mott Relation.} Temperature gradients couple to the distribution function via the real-space gradient operator 
$\bm{r}\cdot\bm{\nabla}_r =\bm{r}\cdot  \bm{\nabla}_r [ T(\bm{r})\partial_T+ \hat{\bm{m}}(\bm{r})\cdot\bm{\nabla}_{\hat{m}}]$
in the Boltzmann equation. In contrast, electric field perturbations only enter the Boltzmann equation through the 
semiclassical equations of motion: $\dot{\bm{r}}\rightarrow \dot{\bm{r}}_E$ and $\dot{\bm{k}}\rightarrow \dot{\bm{k}}_E$ (see Appendix \ref{appD}).  However, even in the presence of all phase space Berry curvatures, the electric field dependent perturbations to the equations of motion can be rewritten such that the electric field dependent part of $\dot{\bm{r}}_E\cdot \bm{\nabla}_r + \dot{\bm{k}}_E\cdot \bm{\nabla}_k$ takes the form $\dot{\bm{r}}\cdot{\bm{E}}\partial_{\widetilde{\mathcal{E}}}$. This allows a simple relationship between the different field dependent perturbations to the distribution function to be established.

The solution to the Boltzmann equation requires inverting the operator $1+\mathds{P}$ with $\mathds{P}_l=\tau(\dot{\bm{r}}_l\cdot \bm{\nabla}_r+\dot{\bm{k}}_l\cdot\bm{\nabla}_k)$. The formal solution can then be written as
\begin{equation}
    g_l=\tau\partial_T f^0[\tilde{\mathcal{E}}_l]\sum_{n=0}^\infty (-\mathds{P}_l)^n\dot{\bm{r}}_l\cdot(-\bm{\nabla}_r T(\bm{r}))
    \label{gfunc}
\end{equation}
\noindent
(see Appendix \ref{appB}).  Similarly in the presence of a constant electric field $\bm{E}=-\bm{\nabla}_r\phi(\bm{r})$ the linear response solution can be written as \cite{verma2022unified}
\begin{equation}
   g_l^{\phi}=\tau e\partial_\varepsilon f^0[\tilde{\mathcal{E}}_l]\sum_{n=0}^\infty (-\mathds{P}_l)^n\dot{\bm{r}_l}\cdot(-\bm{\nabla}_r\phi(\bm{r}))
\end{equation}
\noindent
The charge current $\bm{j}$ deriving from these contributions takes the form $\dot{\bm{r}}_lg_l$ which allows one to determine a Mott relation between $\te{\alpha}^{(2)}$ and $\partial \te{\sigma}^{(2)}/\partial \mu$ where $\mu$ is the chemical potential.  Similarly by generalizing the work of Ref. \cite{xiao2006berry} we find that a Mott relation also holds for $\te{\alpha}^{(1)}$ and $\partial\te{\sigma}^{(1)}/\partial \mu$ in the regime $a\ll L_s$ (see Appendix \ref{appD}).  By adding these two contributions we arrive at the Mott relation for the full response tensors
\begin{equation}
\alpha_{ij}= -\dfrac{\pi^2}{3}\dfrac{k_B^2T}{e}\dfrac{\partial \sigma_{ij}}{\partial \mu}
\label{MottRel}
\end{equation}
\noindent
valid for $k_BT\ll E_F$.

%The Mott relation \eqref{MottRel} is valid to all orders in the small parameters of our theory.  To calculate the anomalous and topological contributions to the thermoelectric conductivity and Nernst signal, we expand the electric conductivity to second order and utilize Eq. \eqref{MottRel}.

The leading order contribution to the anomalous Hall conductivity $\sigma_{H}^A=(\sigma_{xy}^{(1)}-\sigma_{yx}^{(1)})/2$ derives from the anomalous velocity that is proportional to $ \Omega_{k_xk_y}$ and can be written as
\begin{align}
    \sigma^A_{H}&=-\dfrac{e^2}{2\hbar}\sum_{l=\pm} \dfrac{l}{J^2}\int \dfrac{d^3k}{(2\pi)^3}f^0[\varepsilon_l] \,\hat{\bm{z}}\cdot\bigg(\partial_{k_x}\bm{d}_l\times \partial_{k_y}\bm{d}_l\bigg) \nonumber\\
    &=\dfrac{e^2}{2\hbar} \text{Det}(\te{\chi}) \sum_{l=\pm} l\bar{m}_z\dfrac{t\lambda^2}{J^2}\int \dfrac{d^3k}{(2\pi)^3} \mathscr{Q}(\bm{k})\partial_{\varepsilon_l} f^0[\varepsilon_l] 
    \label{sigAH}
\end{align}
\noindent
where we have used an integration by parts to restrict the integration to momenta near the Fermi surface \cite{haldane2004berry} (see Appendix \ref{appE}).
Using the Mott relation, we find that the leading order contributions to $\alpha_H=\alpha_H^A+\alpha_H^T$ are
\begin{align}
\alpha_H^A&=\bigg(\dfrac{k_B e}{\hbar}\bigg)\bigg(\dfrac{\lambda^2k_BT}{J^2t}\bigg)\bigg(\bar{m}_z\dfrac{\text{Det}(\te{\chi})}{2\pi}\bigg)\mathscr{A}(\mu) \nonumber \\
\alpha_H^T&=-\bigg(\dfrac{k_B e}{\hbar}\bigg)\bigg(\dfrac{tk_BT}{(\hbar/\tau)^2}\bigg)\bigg( n_{\text{top}}a^2\bigg)\mathscr{A}(\mu) \nonumber \\
\mathscr{A}(\mu)&=\dfrac{t^2}{24}\dfrac{\partial}{\partial\mu}\bigg(\sum_{l=\pm1} l \int d^3k\, \partial_{\varepsilon_l}f^0[\varepsilon_l]\,\mathscr{Q}(\bm{k})\bigg)
    \label{alpha-and-A}
\end{align}
\noindent
$\mathscr{Q}(\bm{k})$ is defined in \eqref{Q}, and  $\mathscr{A}(\mu)$ is a dimensionless function of $\mu/t$ and $J/t$ and describes the chemical potential or density dependence of $\alpha_H$ (see Appendix \ref{appE}). 

Note that $\alpha_H^T$ and $\alpha_H^A$ derive from very different mechanisms, the former from the real space Berry curvature and the latter from the  momentum space Berry curvature. Nevertheless, both contributions to the thermoelectric conductivity can be shown to be proportional to 
$\mathscr{Q}(\bm{k})$ and thus have the same functional dependence with the chemical potential or density.  

To understand why $\mathscr{Q}(\bm{k})$ appears in both contributions, we note that the totally anti-symmetric part of any rank two tensor must be invariant under rotations about the $\hat{\bm{z}}$-axis and must change sign under vertical mirror planes. $\mathscr{Q}(\bm{k})$ transforms trivially under rotations about the $\hat{\bm{z}}$-axis, and at this level of our perturbation expansion, it is the natural object to construct from one and two momentum space derivatives of $\varepsilon(\bm{k})$. In $\alpha_H^T$, vertical mirror operations flip the sign of $n_{\text{top}}$. In $\alpha_H^A$, it is $\bar{m}_z$ which changes sign under such a transformation, while $\text{Det}(\te{\chi})$ is left invariant.  See Appendix \ref{appE}.

In the regime $\sigma_L\gg \sigma_{ij}$, $i\neq j$, with $\sigma_L=(\sigma_{xx}+\sigma_{yy}+\sigma_{zz})/3$, and using Eq.~\eqref{MottRel} the Nerst signal can be written as \cite{behnia2015}

\begin{equation}
    N=-\dfrac{\pi^2}{3}\dfrac{k_B^2T}{e}\dfrac{\partial \tan(\Theta_H)}{\partial \mu}
    \label{nernst}
\end{equation}

\noindent
where $\tan(\Theta_H)=\sigma_H/\sigma_L$ is the Hall angle.  For simple metals in the presence of an external magnetic field, a Sondheimer cancellation \cite{sondheimer1948theory} can occur whereby the dominant contributions to the Hall and longitudinal conductivities have similar $\mu$-dependences, so that  $\partial\Theta_H/\partial\mu$ is small and $N$ can be highly suppressed. This cancellation can be avoided by an energy-dependent scattering mechanism \cite{esfarjani2021nernst}. However,  even with a constant relaxation time, the anomalous and topological contributions to $N$ avoid Sondheimer cancellation because the Berry curvatures have opposite signs in spin-split bands (Appendix \ref{appF}).

In parallel to Eq.~\eqref{alpha-and-A}, we write the contributions to the Nernst effect $N=N^A+N^T$ as 

\begin{align}
N^A&=\bigg(\frac{k_B}{e}\bigg)\bigg(\dfrac{\lambda^2(\hbar/\tau) k_B T}{J^2 t^2}\bigg)\bigg(\bar{m}_z\dfrac{\text{Det}(\te{\chi})}{2\pi}\bigg)\mathscr{N}(\mu) \nonumber \\
N^T&=-\bigg(\frac{k_B}{e}\bigg)\bigg(\dfrac{ k_B T}{(\hbar/\tau)}\bigg)\bigg(n_{\text{top}} a^2 \bigg)\mathscr{N}(\mu) \nonumber \\
\mathscr{N}(\mu)&=\pi^3 t^3\frac{\partial}{\partial \mu}\frac{\sum_l l \int d^3k \, \mathscr{Q}(k) \,\partial_{\varepsilon_l}f^0[\varepsilon_l]}{\sum_l \int d^3k \, |\bm{\nabla}_k\varepsilon_l|^2 \,\partial_{\varepsilon_l}f^0[\varepsilon_l] }
\label{NernstMu}
 \end{align}
$\mathscr{Q}(\bm{k})$ is defined in \eqref{Q}, and  $\mathscr{N}(\mu)$ is a dimensionless function of $\mu/t$ and $J/t$ and describes the chemical potential or density dependence of $N$.

\begin{figure}[ht]
\includegraphics[width=.45\textwidth]{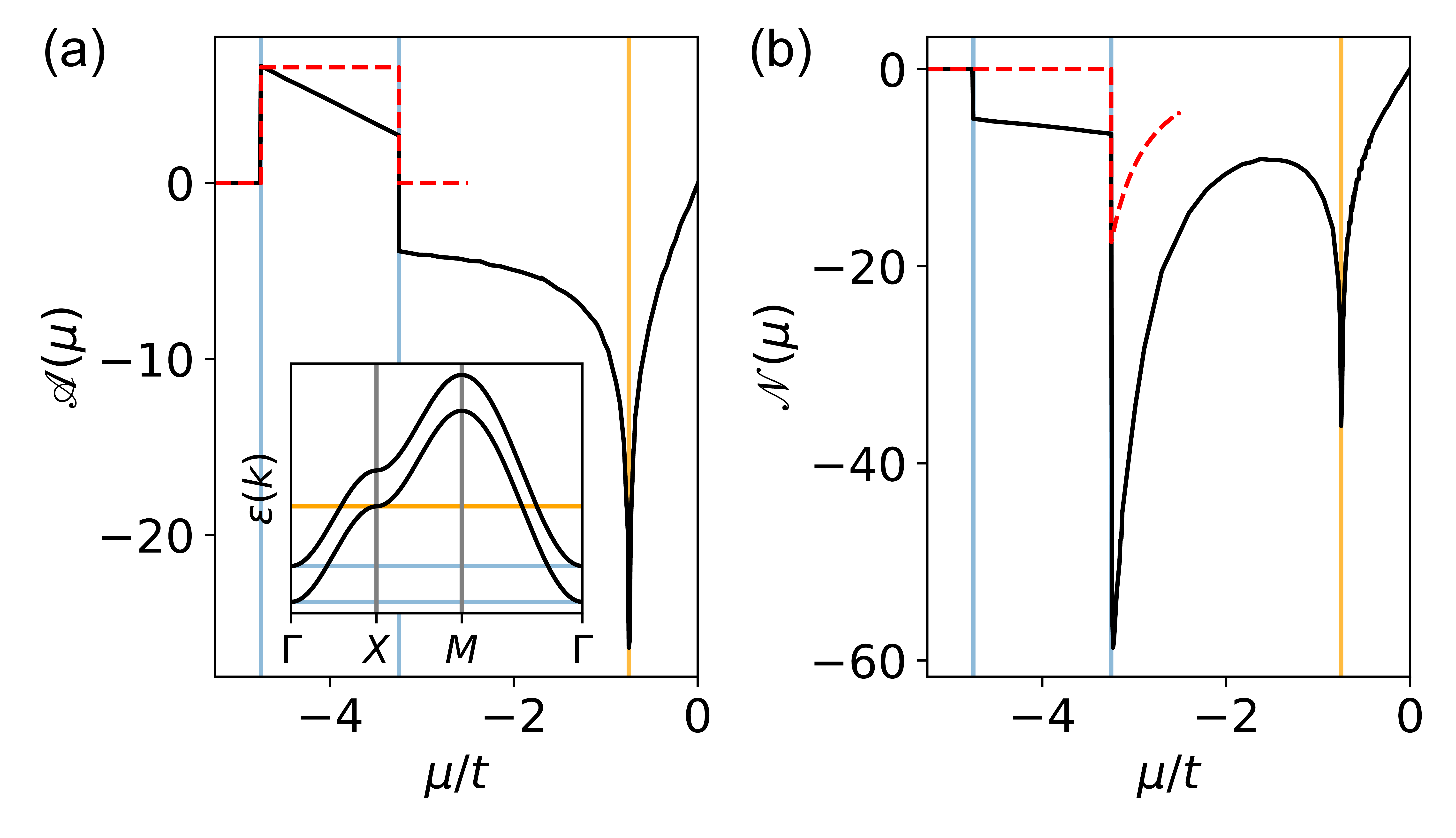}
\caption{{\bf Chemical potential dependence of the thermoelectric conductivity and Nernst signal.} $\mathscr{A}(\mu)$ and $\mathscr{N}(\mu)$ calculated at $T=0$ for the 2D square lattice with $J/t = 3/4$ (black). $\mathscr{N}(\mu)$ and $\mathscr{A}(\mu)$ are odd functions in $\mu/t$ due to particle-hole symmetry in the square lattice. Peaks are observed near the van Hove singularities (gold) and near band edges (blue). The quadratic band approximations for both are plotted near band edges (red dashed lines). See Appendix \ref{appF} for parameter dependence.
}
\label{nernstalphaPlot}
\end{figure}

\medskip
\noindent
\textbf{Model Calculations.}  Given an arbitrary band structure, one can use equations \eqref{alpha-and-A} and \eqref{NernstMu} to compute the thermoelectric conductivity and Nernst signals.  As illustrative examples, we calculate these transport signals in 2D for a system with a parabolic dispersion and for a tight binding model.

%For metallic systems, $k_BT$ is much smaller than electronic energy scales, and it is common in the Mott relation to make the approximation $\partial \sigma_{ij}/\partial \mu \rightarrow \partial \sigma_{ij}/\partial \mu |_{T=0}$.  In the following analysis we will restrict ourselves to the leading order in $T$ behavior of $\te{\alpha}$.

Consider parabolic bands with arbitrary SOC in 2D.  To calculate $\mathscr{A}(\mu)$, we use $\varepsilon_\pm(\bm{k})= \varepsilon_0+ a^2k^2\pm J$ and Eq.~\eqref{alpha-and-A} to find that $\mathscr{A}(\mu)= 2\pi^2/3\,(\Theta[\mu-(\varepsilon_0-J)]-\Theta[\mu-(\varepsilon_0+J)])$ (see dotted lines in Fig. \ref{nernstalphaPlot}(a)). The non-analytic structure in $\mathscr{A}(\mu)$ occurs at the band edges.  For the Nernst signal $\mathscr{N}(\mu)= -4\pi^3 t J/3\mu^2 \, \Theta(\mu-(\varepsilon_0+J))$ and is nonzero only when electron states in both bands are occupied (see Fig. \ref{nernstalphaPlot}(b)).

Fig. \ref{nernstalphaPlot}a shows $\mathscr{A}(\mu)$ and Fig. \ref{nernstalphaPlot}b shows $\mathscr{N}(\mu)$, calculated for nearest neighbor interactions on the two dimensional square lattice: $\varepsilon(\bm{k})=-2t(\cos (k_xa)+\cos (k_ya))$. The quadratic band approximation with $\varepsilon_0= -4t$ is marked by the dashed red lines. The non-analytic jumps in $\mathscr{A}(\mu)$ and $\mathscr{N}(\mu)$ are due to the non-analyticity of the density of states as the chemical potential crosses the band edge.

%For most experimentally relevant temperatures the Mott relation is a poor approximation for systems whose chemical potential is very close to the band edge as $k_BT$ is of order $E_F$.  In these cases, it is more appropriate to employ directly the exact expression in Eq. \eqref{trueAlpha} which shows similar behavior, with the non-analyticity now thermally smeared.

Near the Van Hove singularities, $\mathscr{A}(\mu)$ and $\mathscr{N}(\mu)$ are sharply enhanced as is commonly recognized in signatures of thermoelectric transport \cite{mahan1996best, xia2019high}. In fact, anywhere the density of states changes rapidly with the chemical potential will show an enhancement.

\medskip
\noindent
\textbf{Discussion.}  Through a controlled semiclassical analysis in the regime where $\lambda<J \ll E_F$, $a\ll l\ll L_s$, we have shown that the thermoelectric conductivity is composed of the sum of an anomalous contribution, proportional to the average magnetization, and a topological contribution, proportional to the topological charge density. In addition, we have shown that a Mott relation holds even in the presence of a nonzero topological charge density, thus justifying a commonly held assumption  \cite{shiomi2013topological, hirschberger2020topological, kolincio2021large, mizuta2016large, gayles2018topological,xia2019high}. As a consequence of the Mott relation, the thermoelectric conductivity and Nernst signal are enhanced at points in the band structure in which there is a rapidly changing density of states, such as near van Hove singularities. 

We estimate the order of magnitude of the Nernst signal by approximating Eq. \eqref{NernstMu} with 3D parabolic bands and find $N^T \sim (k_B/e)(k_F \ell)(k_B T/E_F)(n_{\text{top}} a^2)\mathscr{N}(\mu)$. In a skyrmion material with $a/L_s = 1/100$, $n_{\text{top}} a^2 \approx 10^{-3}$. 
We use $k_B T/E_F \approx 10^{-4}$,  and $10 < k_F \ell<100$. For $(t J/\mu^2) \approx 1/3$, we find $\mathscr{N}(\mu) \approx 100$. 
We thus estimate $N^T$ in the range ($10^{-3}$ -- $10^{-2})\,(k_B/e)$ which is equal to 86 - 860 nV/K, similar to what has been 
measured in experiments; see e.g., \cite{hirschberger2020topological}.

We have considered the regime $a\ll \ell \ll L_s$ in the analysis above. The question of how to solve the Boltzmann equation when $a \ll L_s \ll \ell$, the analog of the ``strong field limit’’, is an open question. We have focused here on the intrinsic part of the anomalous thermoelectric response arising from k-space Berry curvature, known to be the dominant contribution to the anomalous Hall response in many materials. The question of how extrinsic effects like skew and side-jump scattering impact the thermoelectric response has not been explored. These are all important questions for future research.

%\cite{bouaziz2021transverse,ishizuka2018spin,akosa2019tuning,zhang2020real}

\medskip
\noindent
\textbf{Acknowledgements.} We thank Nishchhal  Verma for insightful discussion. This work was supported by the NSF Materials Research Science and Engineering Center Grant DMR-2011876. Z.A. was also supported by the Ohio State University President’s Postdoctoral Scholars Program. 

\nocite{}
\bibliography{Bibliography}

\clearpage 

\onecolumngrid

\appendix

\section{Semiclassical Equations of Motion}\label{appA}

Following \cite{xiao2010berry} we can construct a semiclassical theory of Eq. \eqref{eq:ham-gen} by making an expansion of the Hamiltonian about some position $\bm{r}_c$

\begin{align}
    \widehat{H}&\approx\widehat{H}_c+\Delta\widehat{H} \\
    \widehat{H}_c&= -t\sum_{\langle i j\rangle,\sigma}\hat{c}^{\dagger}_{i\sigma}\hat{c}^{\phantom{\dagger}}_{j\sigma} -J \sum_{i,\sigma\sigma'}\hat{c}^{\dagger}_{i\sigma}(\hat{\bm{m}}(\bm{r}_c)\cdot \boldsymbol{\sigma}^{\sigma\sigma'})\hat{c}^{\phantom{\dagger}}_{i\sigma'}+\dfrac{\lambda\hbar}{at} \sum_{\langle ij\rangle,\gamma\delta,\sigma\sigma'}v_\gamma^{ij}\,\hat{c}^{\dagger}_{i\sigma}(\chi^{\phantom{\dagger}}_{\gamma\delta}\sigma_\delta^{\sigma\sigma'}+\dfrac{1}{2}w_\gamma\sigma_z^{\sigma\sigma'})\hat{c}^{\phantom{\dagger}}_{j\sigma'}\\
    \Delta\widehat{H}&=-J\sum_{i,\sigma\sigma'} \hat{c}^{\dagger}_{i\sigma}\bigg((\bm{r}_i-\bm{r}_c)\cdot\bm{\nabla}_{r_c} \hat{\bm{m}}(\bm{r}_c)\cdot\bm{\sigma}^{\sigma\sigma'}\bigg)\hat{c}^{\phantom{\dagger}}_{i\sigma'}
    \label{Hamiltonian}
\end{align}

\noindent
where $\gamma, \delta \in \{x, y\}$ and $\sigma,\sigma'\in \{\uparrow,\downarrow\}$. Here for generality we include a general ising spin-orbit interaction ($\propto\bm{v}^{ij}w_\gamma \sigma_z$).  The semiclassical approximation is valid if $\hat{\bm{m}}(\bm{r})$ is changing slowly in space compared to the other length scales of the problem (i.e. $L_s\gg \ell, a$). 

We can define the semiclassical Bloch Hamiltonian by writing $\widehat{H}_c$ in a semiclassical Bloch basis ($\mathcal{H}_{sc}(\bm{r}_c,\bm{k})\propto e^{-i\bm{k}\cdot\widehat{\bm{r}}} \widehat{H}_c e^{i\bm{k}\cdot\widehat{\bm{r}}}$)

\begin{align}
\mathcal{H}_{sc}(\bm{r}_c,\bm{k})&=\varepsilon(\bm{k})\mathds{1}+ {\bf d}(\bm{r}_c,\bm{k}) \cdot {\boldsymbol\sigma}
    \label{BlochHam}
\end{align}

\noindent
where  $d_\gamma(\boldsymbol\xi) = \sum_{\delta}\dfrac{\lambda}{a\,t}\,\chi^{\phantom{\dagger}}_{\gamma\delta}\partial_{k_\delta}\varepsilon - J \hat{m}_\gamma({\bf r}_c)$; $d_z(\bm{\xi})=\sum_{\gamma}\dfrac{\lambda}{2at}w_{\gamma}\partial_{k_\gamma}\varepsilon-J\hat{m}_z(\bm{r}_c)$ and $\varepsilon(\bm{k})=t\sum\limits_{\langle ij\rangle}e^{i\bm{k}\cdot(\bm{r}_i-\bm{r}_j)}$.  The eigenfunctions $\ket{u_\pm(\bm{r}_c,\bm{k})}$ are the periodic part of the Bloch eigenvectors $\ket{\psi_\pm(\bm{r}_c,\bm{k})}$ and can be determined in terms of the spherical components of $\bm{d}(\bm{r}_c,\bm{k})$, $\theta(\bm{r}_c,\bm{k})$ and $\phi(\bm{r}_c,\bm{k})$.  In the spin-$\hat{\bm{z}}$ basis they can be written as

\begin{align}
    \bm{u}_+(\bm{r}_c,\bm{k})&=\left(
    \begin{array}{c}
    \sin\bigg(\dfrac{\theta(\bm{r}_c,\bm{k})}{2}\bigg)e^{-i\phi(\bm{r}_c,\bm{k})} \\
    -\cos\bigg(\dfrac{\theta(\bm{r}_c,\bm{k})}{2}\bigg)
    \end{array}
    \right)  \nonumber \\
    \bm{u}_-(\bm{r}_c,\bm{k})&=\left(
    \begin{array}{c}
    \cos\bigg(\dfrac{\theta(\bm{r}_c,\bm{k})}{2}\bigg)e^{-i\phi(\bm{r}_c,\bm{k})} \\
    \sin\bigg(\dfrac{\theta(\bm{r}_c,\bm{k})}{2}\bigg)
    \end{array}
    \right)
    \label{eigenstates}
\end{align}

\noindent
and their semiclassical Bloch eigenvalues are 

\begin{equation}
\mathcal{E}_\pm(\bm{r}_c,\bm{k})= \varepsilon(\bm{k})\pm |{\bf d}(\bm{r}_c,\bm{k})|
\label{energies}
\end{equation}

\subsection{Wave Packet Dynamics}

We can construct wave packets from semiclassical Bloch states $\ket{\psi_\pm(\bm{r}_c,\bm{k})}$ via

\begin{equation}
    \ket{W_\pm(\bm{r}_c,\bm{k}_c)}=\sum_{\bm{k}}\gamma_\pm(\bm{r}_c,\bm{k})\ket{\psi_\pm(\bm{r}_c,\bm{k})}
    \label{wavepacket}
\end{equation}

\noindent
Here $\gamma_\pm(\bm{r}_c,\bm{k})$ is chosen such that the wave packet is strongly peaked at $\bm{r}_c$ and $\bm{k}_c$.

\begin{align}
    \bra{W_\pm(\bm{r}_c,\bm{k}_c)}\hat{\bm{r}}\ket{W_\pm(\bm{r}_c,\bm{k}_c)}&=\bm{r}_c \\
    \bra{W_\pm(\bm{r}_c,\bm{k}_c)}\hat{\bm{p}}\ket{W_\pm(\bm{r}_c,\bm{k}_c)}&=\hbar\bm{k}_c \\
\end{align}

\noindent
To find the time evolution of $\bm{r}_c$ and $\bm{k}_c$ we construct the semiclassical Lagrangian

\begin{equation}
    L_\pm(\bm{r}_c,\bm{k}_c)=\bra{W_\pm(\bm{r}_c,\bm{k}_c)}i\hbar\dfrac{\partial}{\partial t}-\widehat{H}\ket{W_\pm(\bm{r}_c,\bm{k}_c)}
    \label{lagrange}
\end{equation}

\noindent
and find the Euler-Lagrange equations of motion for $\bm{r}_c$ and $\bm{k}_c$.  These equations can be written in compact form as (see \cite{xiao2010berry}) \begin{equation}
    \sum_\beta \Gamma^\pm_{\alpha\beta}(\bm{\xi})\dot{\xi}_\beta=\dfrac{1}{\hbar}\nabla_{\xi_\alpha}\widetilde{\mathcal{E}}_\pm(\boldsymbol\xi)
    \label{EOM}
\end{equation}

\noindent
Here $\bm{\xi}=(r_x,r_y,r_z,k_x,k_y,k_z)$, $\dot{\bm{\xi}}=\partial_t\bm{\xi}$, and we drop the $c$ label on $\bm{r}_c$ and $\bm{k}_c$ for convenience.  The energy functional that appear in the semiclassical equations is

\begin{align}
\widetilde{\mathcal{E}}_\pm(\boldsymbol\xi)&=\mathcal{E}_\pm(\bm{\xi})+\Delta\mathcal{E}_\pm(\bm{\xi}) \nonumber \\
\Delta\mathcal{E}_\pm(\bm{\xi})&=\bra{W_\pm(\bm{\xi})}\Delta\widehat{H}\ket{W_\pm(\bm{\xi})} \nonumber \\
&=-\sum_i\text{Im}\bigg((\partial_{r_i}\bra{u_\pm(\bm{\xi})})\bigg(\mathcal{E}_\pm(\bm{\xi})-\mathcal{H}_{sc}(\bm{\xi})\bigg)(\partial_{k_i}\ket{u_\pm(\bm{\xi}})\bigg)
\end{align}

\noindent
and $\Gamma^\pm(\bm{\xi})$ is a rank two totally antisymmetric tensor of dimension four.

\begin{equation}
\Gamma^\pm(\bm{\xi}) = \left[ \begin{pmatrix}
0 & \Omega^\pm_{r_xr_y} & \Omega^\pm_{r_xr_z} & \Omega^\pm_{r_xk_x} & \Omega^\pm_{ r_xk_y } & \Omega^\pm_{ r_xk_z } \\
-\Omega^\pm_{r_x r_y} & 0 & \Omega^\pm_{r_yr_z} & \Omega^\pm_{ r_yk_x} & \Omega^\pm_{ r_y k_y } & \Omega^\pm_{ r_y k_z }  \\
-\Omega^\pm_{r_xr_z} & -\Omega^\pm_{r_yr_z} & 0 & \Omega^\pm_{r_zk_x} & \Omega^\pm_{r_zk_y} & \Omega^\pm_{r_zk_z} \\
-\Omega^\pm_{r_x k_x} & -\Omega^\pm_{ r_y k_x} & -\Omega^\pm_{ r_z k_x} &  0 & \Omega^\pm_{ k_x k_y } & \Omega^\pm_{ k_x k_z }  \\
- \Omega^\pm_{ r_x k_y } & - \Omega^\pm_{r_y k_y} & -\Omega^\pm_{r_zk_y} & - \Omega^\pm_{ k_x k_y } & 0 & \Omega^\pm_{ k_y k_z } \\
-\Omega^\pm_{r_xk_z} & -\Omega^\pm_{r_yk_z} & - \Omega^\pm_{r_zk_z} & -\Omega^\pm_{k_xk_z} & -\Omega^\pm_{k_yk_z} & 0
\end{pmatrix} - \begin{pmatrix}
0 & 0 & 0 & 1 & 0 & 0 \\
0 & 0 & 0 & 0 & 1 & 0\\
0 & 0 & 0 & 0 & 0 & 1\\
-1 & 0 & 0 & 0 & 0 & 0\\
0 & -1 & 0 & 0 & 0 & 0 \\
0 & 0 & -1 & 0 & 0 & 0\
\end{pmatrix}
\right]
\end{equation}

\noindent
where $\Omega^\pm_{\xi_\alpha\xi_\beta}$ are the components of generalized curvatures in the expanded phase space spanned by $\bm{r}$ and $\bm{k}$

\begin{align}
    \Omega^\pm_{\xi_\alpha\xi_\beta}&=\partial_{\xi_\alpha}\mathcal{A}^\pm_{\xi_\beta}-\partial_{\xi_\beta}\mathcal{A}^\pm_{\xi_\alpha} \nonumber\\
    \mathcal{A}^\pm_{\xi_\alpha}&=i\bra{u_\pm(\bm{\xi})}\partial_{\xi_\alpha}\ket{u_\pm(\bm{\xi})}
    \label{concurv}
\end{align}

\noindent
Note that for simplicity we suppress the dependence of the curvatures on $\bm{\xi}$ (i.e. $\Omega_{\xi_\alpha\xi_\beta}(\bm{\xi})\rightarrow \Omega_{\xi_\alpha\xi_\beta}$) and we suppress the band index.  We may solve Eq. \eqref{EOM}  for systems with a Hamiltonian of the form in Eq. \eqref{eq:ham-gen}. In the following analysis, we assume that $\hat{\bm{m}}(\bm{r})$ is independent of $r_z$ as would be the case, for example, in a skyrmion tubes phase. This restriction causes all Berry curvatures involving $r_z$ or $k_z$ derivatives to vanish. Subsequently, the equations of motion for the phase space variables $\dot{\bm{\xi}}=(\dot{r}_x,\dot{r}_y,\dot{r}_z,\dot{k}_x,\dot{k}_y,\dot{k}_z)$ are

\begin{align}
    \dot{r}_{\gamma}(\bm{\xi})&=\dfrac{1}{\mathcal{D}(\bm{\xi})}\sum_{\delta}\bigg([\te{\Omega}_{kk}]_{\gamma\delta} \dfrac{\partial_{r^{\phantom{\dagger}}_\delta} \widetilde{\mathcal{E}}(\boldsymbol\xi)}{\hbar}+[(\mathds{1}+\te{\Omega}_{rk})]_{\gamma\delta}\dfrac{\partial_{k^{\phantom{\dagger}}_\delta}\widetilde{\mathcal{E}}(\boldsymbol\xi)}{\hbar}\bigg) \nonumber \\
    \dot{k}_\gamma(\bm{\xi})&=\dfrac{1}{\mathcal{D}(\bm{\xi})}\sum_{\delta}\bigg([\te{\Omega}_{rr}]_{\gamma\delta} \dfrac{ \partial_{k^{\phantom{\dagger}}_\delta} \widetilde{\mathcal{E}}(\boldsymbol\xi)}{\hbar}-[(\mathds{1}+\te{\Omega}_{rk}^T)]_{\gamma\delta}\dfrac{\partial_{r^{\phantom{\dagger}}_\delta} \widetilde{\mathcal{E}}(\boldsymbol\xi)}{\hbar}\bigg)
    \\
    \dot{r}_z&=\dfrac{1}{\mathcal{D}(\bm{\xi})} \dfrac{\partial_{k_z}\tilde{\mathcal{E}}(\bm{\xi})}{\hbar},\,\,\,\dot{k}_z=0
    \label{xidot}
\end{align}

\noindent
where $\gamma,\delta\in\{x,y\}$.  The superscript $T$ stands for transpose and the two dimensional generalized curvature matrices are

\begin{align}
    &\Gamma^{-1} (\bm{\xi})=\dfrac{1}{\mathcal{D} (\bm{\xi})}\begin{pmatrix}
\te{\Omega} _{kk} & \mathds{1}+\te{\Omega} _{rk} \\
-(\mathds{1}+\te{\Omega} _{rk}^T) & \te{\Omega} _{rr} \\
\end{pmatrix} &  \\
&\te{\Omega} _{kk}=\begin{pmatrix}
0 & \Omega _{k_xk_y} \\
-\Omega _{k_xk_y} & 0 
\end{pmatrix} \,\,,\,\,\,\,
\te{\Omega} _{rr}=\begin{pmatrix}
0 & \Omega _{r_xr_y} \\
-\Omega _{r_xr_y} & 0 
\end{pmatrix}  \,\,,\,\,\,\,
{\te{\Omega}} _{rk}=\begin{pmatrix}
-\Omega _{r_yk_y} & \Omega _{r_yk_x} \\
\Omega _{r_xk_y} & -\Omega _{r_xk_x}
\end{pmatrix} \\
&\mathcal{D} (\bm{\xi})=\sqrt{\det(\Gamma (\bm{\xi}))}=|1-\Omega _{r_xk_x}-\Omega _{r_yk_y}-\Omega _{r_xr_y}\Omega _{k_xk_y}
-\Omega _{r_xk_y}\Omega _{r_yk_x}+\Omega _{r_xk_x}\Omega _{r_yk_y}|
\end{align}

\section{Thermal Perturbations and the Boltzmann Equation}\label{appB}

To calculate the Nernst effect we must calculate the charge current to first order in spatial derivatives of the temperature.  Contributions proportional to the scattering time $\tau$ derive from temperature gradient induced corrections to the electronic distribution function that can be determined by solving the Boltzmann equation. 

\begin{equation}
    \dfrac{d}{dt}f(\bm{\xi})=\partial_t f(\bm{\xi})+\dot{\bm{r}}(\bm{\xi})\cdot\bm{\nabla}_rf(\bm{\xi})+\dot{\bm{k}}(\bm{\xi})\cdot\bm{\nabla}_k f(\bm{\xi})
\end{equation}

\noindent
Note that there is no explicit time-dependence in $\bm{H}$ such that $\partial_tf(\bm{\xi})=0$.  For simplicity in what follows we will suppress the band index.  Here we solve this equation under the relaxation time approximation

\begin{equation}
    \dfrac{d}{dt}f(\bm{\xi})\approx -\dfrac{f(\bm{\xi})-f^0(\bm{\xi})}{\tau}
\end{equation}

\noindent
with relaxation time $\tau$.  Here $f^0(\bm{\xi})$ is the equilibrium distribution function

\begin{equation}
    f^0(\bm{\xi})=\dfrac{1}{e^{(\varepsilon(\bm{\xi})-\mu)/k_B T}+1}
\end{equation}

\noindent 
In the presence of temperature gradients we have

\begin{equation}
    -\dfrac{f(\bm{\xi})-f^0(\bm{\xi})}{\tau}=\dot{\bm{r}}(\bm{\xi})\cdot (\bm{\nabla}_r \hat{\bm{m}}(\bm{r})\cdot \bm{\nabla}_m f(\bm{\xi})+\bm{\nabla}_r T(\bm{r})\partial_T f(\bm{\xi}))+\dot{\bm{k}}(\bm{\xi})\cdot\bm{\nabla}_k f(\bm{\xi})
    \label{boltzmann}
\end{equation}

\noindent
where we have used

\begin{equation}
    \bm{\nabla}_rf(\bm{\xi})=\bm{\nabla}_r \hat{\bm{m}}(\bm{r})\cdot \bm{\nabla}_m f(\bm{\xi})+\bm{\nabla}_r T(\bm{r})\partial_T f(\bm{\xi})
\end{equation}

\noindent
with

\begin{equation}
    \bm{\nabla}_m=\bigg(\dfrac{\partial}{\partial m_{r_x}},\dfrac{\partial}{\partial m_{r_y}},\dfrac{\partial}{\partial m_{r_z}}\bigg)
\end{equation}

Without the statistical drive induced by the temperature gradient the distribution function is at equilibrium and the Boltzmann equation dictates

\begin{equation}
    0=\dot{\bm{r}}(\bm{\xi})\cdot\bm{\nabla}_rf^0(\bm{\xi})+\dot{\bm{k}}(\bm{\xi})\cdot\bm{\nabla}_k f^0(\bm{\xi})
    \label{eqID}
\end{equation}

\noindent
Using the chain rule and the equation of motion we show that this is satisfied if $f^0(\bm{\xi})$ is a function of $\varepsilon(\bm{\xi})$ alone (i.e. $f^0(\bm{\xi})=f^0[\widetilde{\mathcal{E}}(\boldsymbol\xi)]$):

\begin{align}
    0&=\bigg(\dot{\bm{r}}(\bm{\xi})\cdot\bm{\nabla}_r\widetilde{\mathcal{E}}(\boldsymbol\xi)+\dot{\bm{k}}(\bm{\xi})\cdot\bm{\nabla}_k\widetilde{\mathcal{E}}(\boldsymbol\xi)\bigg) \partial_{\widetilde{\mathcal{E}}} f^0[\widetilde{\mathcal{E}}(\boldsymbol\xi)] \nonumber \\
    &=\sum_{\delta\gamma}\bigg[\bigg([\te{\Omega}_{kk}]_{\gamma\delta} \dfrac{\partial_{r^{\phantom{\dagger}}_\delta} \widetilde{\mathcal{E}}(\boldsymbol\xi)}{\hbar}+[(\mathds{1}+\te{\Omega}_{rk})]_{\gamma\delta}\dfrac{\partial_{k^{\phantom{\dagger}}_\delta}\widetilde{\mathcal{E}}(\boldsymbol\xi)}{\hbar}\bigg)\partial_{r^{\phantom{\dagger}}_{\gamma}}\widetilde{\mathcal{E}}(\boldsymbol\xi) \nonumber \\
    &+\bigg([\te{\Omega}_{rr}]_{\gamma\delta} \dfrac{ \partial_{k^{\phantom{\dagger}}_\delta} \widetilde{\mathcal{E}}(\boldsymbol\xi)}{\hbar}-[(\mathds{1}+\te{\Omega}_{rk}^T)]_{\gamma\delta}\dfrac{\partial_{r^{\phantom{\dagger}}_\delta} \widetilde{\mathcal{E}}(\boldsymbol\xi)}{\hbar}\bigg) \partial_{k^{\phantom{\dagger}}_\gamma}\widetilde{\mathcal{E}}(\boldsymbol\xi)\bigg]\dfrac{\partial_{\widetilde{\mathcal{E}}} f^0[\widetilde{\mathcal{E}}(\boldsymbol\xi)]}{\mathcal{D}(\varepsilon)} \nonumber \\
    &=\sum_{\delta\gamma}\bigg([\te{\Omega}_{rk}]_{\gamma\delta}\dfrac{\partial_{k^{\phantom{\dagger}}_\delta}\widetilde{\mathcal{E}}(\boldsymbol\xi)}{\hbar}\partial_{r^{\phantom{\dagger}}_\gamma}\widetilde{\mathcal{E}}(\boldsymbol\xi)-[\te{\Omega}_{rk}^T]_{\delta\gamma}\dfrac{\partial_{r^{\phantom{\dagger}}_\delta} \widetilde{\mathcal{E}}(\boldsymbol\xi)}{\hbar}\partial_{k^{\phantom{\dagger}}_\gamma}\widetilde{\mathcal{E}}(\boldsymbol\xi)\bigg)\dfrac{\partial_{\widetilde{\mathcal{E}}} f^0[\widetilde{\mathcal{E}}(\boldsymbol\xi)]}{\mathcal{D}(\varepsilon)} \nonumber \\
    &=0
\end{align}

\noindent
Note that $\te{\Omega}_{kk}$, $\te{\Omega}_{rr}$, and $\te{\Omega}_{rk}$ are totally anti-symmetric matrices such that 

\begin{equation}
(\te{\Omega} \bm{v})\cdot \bm{v}=\sum_{ij}\bm{v}_i\te{\Omega}_{ij} \bm{v}_j=0
\end{equation}

\noindent
for all $\bm{v}$.

\subsection{First Order in Temperature Gradients}

To determine the distribution function to first order in temperature gradients we may make a power series expansion of $f(\bm{\xi})$ in powers of the temperature gradients.  Here we only are interested to first order such that we may write

\begin{equation}
    f(\bm{\xi})\approx f^0[\widetilde{\mathcal{E}}(\boldsymbol\xi)]+g(\bm{\xi})
\end{equation}

\noindent
where $g(\bm{\xi})$ is first order in temperature gradients (i.e. $g(\bm{\xi})\propto \bm{\nabla}_r T(\bm{r})$).  Substitution into Eq. \eqref{boltzmann} and equating terms that are first order in temperature gradients we have

\begin{align}
  & \dfrac{ -g(\bm{\xi})}{\tau}=\dot{\bm{r}}(\bm{\xi})\cdot (\bm{\nabla}_r \hat{\bm{m}}(\bm{r})\cdot \bm{\nabla}_m g(\bm{\xi})+\bm{\nabla}_r T(\bm{r})\partial_T f^0[\widetilde{\mathcal{E}}(\boldsymbol\xi)])+\dot{\bm{k}}(\bm{\xi})\cdot\bm{\nabla}_k g^T(\bm{\xi}) \nonumber \\
  \implies & \bigg(\tau(\dot{\bm{r}}(\bm{\xi})\cdot \bm{\nabla}_r \hat{\bm{m}}(\bm{r})\cdot \bm{\nabla}_m+\dot{\bm{k}}(\bm{\xi})\cdot\bm{\nabla}_k)+1\bigg) g(\bm{\xi})=-\tau\dot{\bm{r}}\cdot\bm{\nabla}_r T(\bm{r})\partial_T f^0[\widetilde{\mathcal{E}}(\boldsymbol\xi)]
\end{align}

\noindent
This is in the form

\begin{equation}
    \bigg(1+\mathds{P}\bigg)g(\bm{\xi})=-\tau\dot{\bm{r}}(\bm{\xi})\cdot\bm{\nabla}_r T(\bm{r})\partial_T f^0[\widetilde{\mathcal{E}}(\boldsymbol\xi))]
\end{equation}

\noindent
with $\mathds{P}=\tau(\dot{\bm{r}}(\bm{\xi})\cdot \bm{\nabla}_r \hat{\bm{m}}(\bm{r})\cdot \bm{\nabla}_m+\dot{\bm{k}}(\bm{\xi})\cdot\bm{\nabla}_k)$. Solving for $g^T(\bm{\xi})$ we have

\begin{equation}
    g(\bm{\xi})=-\tau\sum_{n=0}^\infty (-1)^n(\mathds{P})^n\bigg(\dot{\bm{r}}(\bm{\xi})\cdot\bm{\nabla}_r T(\bm{r})\partial_T f^0[\widetilde{\mathcal{E}}(\boldsymbol\xi)]\bigg)
\end{equation}

\noindent
Note here we consider a constant temperature gradients (i.e. $\bm{\nabla}_r T(\bm{r})= \text{constant}$) such that $\mathds{P}$ acts only acts on $\dot{\bm{r}}(\bm{\xi})$ and $\partial_T f^0[\widetilde{\mathcal{E}}(\boldsymbol\xi)]$.  This allows us to write $\bm{\nabla}_r \hat{\bm{m}}(\bm{r})\cdot \bm{\nabla}_m$ as $\bm{\nabla}_r$.  With this simplification we have

\begin{equation}
    g(\bm{\xi})=-\tau\sum_{n=0}^\infty (-\tau)^n(\dot{\bm{r}}(\bm{\xi})\cdot\bm{\nabla}_r+\dot{\bm{k}}(\bm{\xi})\cdot\bm{\nabla}_k)^n\bigg(\dot{\bm{r}}(\bm{\xi})\cdot\bm{\nabla}_r T(\bm{r})\partial_T f^0[\widetilde{\mathcal{E}}(\boldsymbol\xi)]\bigg)
\end{equation}

\noindent
The operator in the second parenthesis acts on the functions $\dot{\bm{r}}(\bm{\xi})$ and $\partial_T f^0[\widetilde{\mathcal{E}}(\boldsymbol\xi)]$.  However Eq. \eqref{eqID} constrains the action of the operator on $\partial_T f^0[\widetilde{\mathcal{E}}(\boldsymbol\xi)]$ to vanish.  This occurs at all orders (i.e. $\mathbb{P}^nf^0[\widetilde{\mathcal{E}}(\boldsymbol\xi)]=0$).  We may rewrite our expression for $g(\bm{\xi})$ as

\begin{equation}
    g(\bm{\xi})=-\tau\partial_T f^0[\widetilde{\mathcal{E}}(\boldsymbol\xi)]\sum_{n=0}^\infty (-\tau)^n(\dot{\bm{r}}(\bm{\xi})\cdot\bm{\nabla}_r+\dot{\bm{k}}(\bm{\xi})\cdot\bm{\nabla}_k)^n\bigg(\dot{\bm{r}}(\bm{\xi})\cdot\bm{\nabla}_r T(\bm{r})\bigg)
    \label{gfunc}
\end{equation}

\subsection{Scaling of Terms in Perturbed Distribution Function}

We have: 

\begin{equation}
\tau(\dot{\bm{r}}(\bm{\xi})\cdot \bm{\nabla}_r +\dot{\bm{k}}(\bm{\xi})\cdot\bm{\nabla}_k)
\approx \tau \bigg(\dfrac{\bm{\nabla}_k\mathcal{E}(\bm{\xi})}{\hbar}\cdot\bm{\nabla}_r-\dfrac{\bm{\nabla}_r\mathcal{E}(\bm{\xi})}{\hbar}\cdot\bm{\nabla}_k\bigg)
\end{equation}

\noindent
Where we have used $\widetilde{\mathcal{E}}(\boldsymbol\xi)\approx \mathcal{E}(\bm{\xi})$. These terms multiply $\partial_T f^0[\widetilde{\mathcal{E}}(\boldsymbol\xi)]$ in $g(\bm{\xi})$ and will thus contribute where  $\partial_T f^0[\widetilde{\mathcal{E}}(\boldsymbol\xi)]$ is large, which is in a range of $k_B T$ around the Fermi energy.  Therefore we have 

\begin{align}
    \tau \bigg(\dfrac{\bm{\nabla}_k\mathcal{E}(\bm{\xi})}{\hbar}\cdot\bm{\nabla}_r-\dfrac{\bm{\nabla}_r\mathcal{E}(\bm{\xi})}{\hbar}\cdot\bm{\nabla}_k\bigg) \sim \tau \bigg(\bm{v}_F\cdot \bm{\nabla}_r+\dfrac{\lambda}{\hbar L_s}\partial_{k}\bigg)\sim \dfrac{\ell}{L_s}
    \label{scaling1}
\end{align}

\noindent
where we note that $J\sim \mathcal{E}_F$ and at most $\partial_{r_i}\sim 1/L_s$ and $\partial_{k_i}\sim a$.  This statement relies heavily on the condition $\lambda \ll J$ and the action of the operator on $\dot{\bm{r}}(\bm{\xi})$ (e.g. the first term in Eq. \eqref{scaling1} actually scales as $(\ell/L_s) (\lambda/\mathcal{E}_F)$). This shows that at leading order $\tau(\dot{\bm{r}}(\bm{\xi})\cdot \bm{\nabla}_r +\dot{\bm{k}}(\bm{\xi})\cdot\bm{\nabla}_k)$ scales as $\ell/L_s$, which in the regime of interest is much less than one.  We thus only need to consider the first few $n$ in the sum of Eq. \eqref{gfunc}.

\section{Contributions to the Thermoelectric Tensor}\label{appC}

\noindent
Contributions to the thermoelectric tensor $\te{\alpha}^{(E)}$ derive from the second term in Eq. \eqref{curr} as the temperature gradient induces $\tau$ dependent corrections to the distribution function.

\begin{align}
    \bm{j}^{(2)}_{\text{tr}}=-e\sum_{l=\pm}\int \dfrac{d^6\xi}{V(2\pi)^3} \mathcal{D}_l(\bm{\xi})\dot{\bm{r}}_l(\bm{\xi})f_l(\bm{\xi})
    \label{currT}
\end{align}

\noindent
to first order in temperature gradients we may write

\begin{align}
    \bm{j}^{(2)}_{\text{tr}}&\approx-e\sum_{l=\pm}\int \dfrac{d^6\xi}{V(2\pi)^3} \mathcal{D}_l(\bm{\xi})\dot{\bm{r}}_l(\bm{\xi})g^T_l(\bm{\xi}) \nonumber \\
    &=e\tau\sum_{l=\pm}\int \dfrac{d^6\xi}{V(2\pi)^3}\mathcal{D}_l(\bm{\xi}) \dot{\bm{r}}_l(\bm{\xi})\partial_Tf^0[\widetilde{\mathcal{E}}_l(\boldsymbol\xi)]\bigg(\sum_{n=0}^\infty (-1)^n(\mathds{P}_l)^n\bigg)\dot{\bm{r}}(\bm{\xi})\cdot\bm{\nabla}_r T(\bm{r})
    \label{currE}
\end{align}

\noindent

such that we may write the thermoelectric conductivity as

\begin{equation}
    \alpha^{(2)}_{ij}=-e\tau\sum_{l=\pm}\int \dfrac{d^6\xi}{V(2\pi)^3}\mathcal{D}_l(\bm{\xi})\partial_Tf^0[\widetilde{\mathcal{E}}_l(\boldsymbol\xi)] \bigg[\dot{r}^i_l(\bm{\xi})\bigg(\sum_{n=0}^\infty (-1)^n(\mathds{P}_l)^n \bigg)\dot{r}^j_l(\bm{\xi})\bigg]
\end{equation}

\subsection{First Order in $\tau$}

To find contributions to the thermoelectric tensor first order in $\tau$ we may substitute $g^{(1)}(\bm{\xi})$ into Eq. \eqref{currE}.  This gives

\begin{align}
    j^{(2,1)}_{\text{tr}}=e\tau\sum_{l=\pm}\int \dfrac{d^6\xi}{V(2\pi)^3} \mathcal{D}_l(\bm{\xi})\dot{\bm{r}}_l(\bm{\xi})\partial_T f^0[\widetilde{\mathcal{E}}(\boldsymbol\xi)]\bigg( \dot{\bm{r}}_l(\bm{\xi})\cdot\bm{\nabla}_r T(\bm{r})\bigg)
    \label{currE1}
\end{align}

\noindent
Extracting the thermoelectric tensor we have

\begin{align}
    \alpha_{ij}^{(2,1)}=-e\tau\sum_{l=\pm}\int \dfrac{d^6\xi}{V(2\pi)^3} \mathcal{D}_l(\bm{\xi}) \partial_T f^0[\widetilde{\mathcal{E}}_l(\boldsymbol\xi)] \,(\dot{\bm{r}}_l(\bm{\xi})\cdot \hat{\bm{i}}) (\dot{\bm{r}}_l(\bm{\xi})\cdot \hat{\bm{j}})
    \label{alphaE1}
\end{align}

\noindent
At this order in $\tau$, the tensor $\alpha_{ij}^{(E,1)}$ is symmetric such that the transverse contribution to $\alpha^{(E,1)}$ vanishes: $\alpha_T^{(E,1)}=1/2\bigg(\alpha_{xy}^{(E,1)}-\alpha_{yx}^{(E,1)}\bigg)=0$.  The leading order longitudinal contribution can be written as

\begin{equation}
\alpha_{L}=-\dfrac{e\tau}{3\hbar^2} \int_\xi\,|\bm{\nabla}_k\varepsilon_l|^2\,\partial_Tf^0[\varepsilon_l]
\end{equation}

\subsection{Second Order in $\tau$}

We have already found the leading order non-vanishing longitudinal  component of the thermoelectric tensor and therefore will focus now on the transverse antisymmetric component of the thermoelectric tensor.  Using Eq. \eqref{currT} the charge current to second order in $\tau$ is 

\begin{equation}
    j^{(2,2)}=-e\tau^2\sum_{l=\pm}\int \dfrac{d^6\xi}{V(2\pi)^3} \mathcal{D}_l(\bm{\xi})\dot{\bm{r}}_l(\bm{\xi})\partial_T f^0[\varepsilon_l(\bm{\xi})] \bigg(\dot{\bm{r}}_l(\bm{\xi})\cdot\bm{\nabla}_r+\dot{\bm{k}}_l(\bm{\xi})\cdot\bm{\nabla}_k\bigg)\bigg( \dot{\bm{r}}_l(\bm{\xi})\cdot\bm{\nabla}_r T(\bm{r})\bigg)
\end{equation}

\noindent
Extracting the thermoelectric tensor we have

\begin{equation}
   \alpha_{ij}^{(2,2)}= e\tau^2\sum_{l=\pm}\int \dfrac{d^6\xi}{V(2\pi)^3} \mathcal{D}_l(\bm{\xi})(\dot{\bm{r}}_l(\bm{\xi})\cdot \hat{\bm{i}})\partial_T f^0[\varepsilon_l(\bm{\xi})] \bigg(\dot{\bm{r}}_l(\bm{\xi})\cdot\bm{\nabla}_r+\dot{\bm{k}}_l(\bm{\xi})\cdot\bm{\nabla}_k\bigg) (\dot{\bm{r}}_l(\bm{\xi})\cdot \hat{\bm{j}})
   \label{aE2}
\end{equation}

\noindent
We now want to find the largest terms in $a/L_s$ and $\lambda/\mathcal{E}_F$.  We start by looking at leading order in $a/L_s$.

\subsubsection{First order in $a/L_s$}

We note that nominally $\dot{\bm{r}}_l(\bm{\xi})\cdot\bm{\nabla}_r\sim \dfrac{\bm{\nabla}_k\mathcal{E}_l(\bm{\xi})}{\hbar}\cdot\bm{\nabla}_r\sim a/L_s$ and at leading order we have $\dot{\bm{k}}_l(\bm{\xi})\cdot\bm{\nabla}_k\sim-\dfrac{\bm{\nabla}_r \mathcal{E}_l(\bm{\xi})}{\hbar}\cdot \bm{\nabla}_k\sim a/L_s$.  Thus to leading order in $a/L_s$ we have

\begin{equation}
    \alpha_{ij}^{(2,2)}\approx e\tau^2\sum_{l=\pm}\int \dfrac{d^6\xi}{V(2\pi)^3}\,\partial_T f^0[\mathcal{E}_l(\bm{\xi})] \bigg(\dfrac{\partial_{k_i}\mathcal{E}_l(\bm{\xi})}{\hbar}\bigg) \bigg(\dfrac{\bm{\nabla}_k\mathcal{E}_l(\bm{\xi})}{\hbar}\cdot\bm{\nabla}_r-\dfrac{\bm{\nabla}_r \mathcal{E}_l(\bm{\xi})}{\hbar}\cdot\bm{\nabla}_k\bigg) \bigg(\dfrac{\partial_{k_j}\mathcal{E}_l(\bm{\xi})}{\hbar}\bigg)
    \label{aE2app}
\end{equation}

\noindent
To find the largest contribution to the above we can now expand this to first order in $\lambda/\mathcal{E}_F$.  This can be done by first expanding $\mathcal{E}_l(\bm{\xi})$ to first order in $\lambda$.

\begin{equation}
    \mathcal{E}_\pm(\bm{\xi})\approx \varepsilon_\pm(\bm{k}) \pm\lambda a/t \sum_{\substack{i=x,y,z \\ j= x,y}} m_i(\bm{r})(\chi_{ij}+\delta_{zj}w_j)\partial_{k_j}\varepsilon(\bm{k})
\end{equation}

\noindent
Thus any spatial gradient will be nominally order $\lambda$ and we see the order $(\lambda/\mathcal{E}_F)^0$ term vanishes in Eq. \eqref{aE2app}.  Thus for all other gradients we may substitute in the $\lambda$ independent part and we may use $f_l^0[\mathcal{E}(\bm{\xi})]\approx f_l^0[\mathcal{E}(k)]$ in the above.  We find

\begin{align}
\dfrac{1}{2}\bigg[\bigg(\dfrac{\partial_{k_x}\mathcal{E}_\pm(\bm{\xi})}{\hbar}\bigg) \bigg(\dfrac{\bm{\nabla}_k\mathcal{E}_\pm(\bm{\xi})}{\hbar}\cdot\bm{\nabla}_r-\dfrac{\bm{\nabla}_r \mathcal{E}_\pm(\bm{\xi})}{\hbar}\cdot\bm{\nabla}_k\bigg) \bigg(\dfrac{\partial_{k_y}\mathcal{E}_\pm(\bm{\xi})}{\hbar}\bigg)-(x\leftrightarrow y)\bigg] \\ \approx \pm \dfrac{at^2\lambda}{2\hbar^3}\mathscr{Q}(\bm{k})\bigg(\partial_{r_y}\bm{m}(\bm{r})\cdot\te{\chi}\cdot\bm{\hat{x}}-\partial_{r_x}\bm{m}(\bm{r})\cdot\te{\chi}\cdot\bm{\hat{y}}\bigg)
\end{align}

\noindent
Substitution into Eq. \eqref{aE2app} gives

\begin{align}
    \delta\alpha_H\equiv \dfrac{1}{2}\bigg( \alpha^{(2,2)}_{xy}-\alpha^{(2,2)}_{yx}\bigg)&=\dfrac{e\tau^2t^2\lambda}{\hbar^3}\bigg[\sum_l\,l\,\int \dfrac{d^3r}{2V}\,a\,\bigg(\partial_{r_y}\bm{m}(\bm{r})\cdot\te{\chi}\cdot\bm{\hat{x}}-\partial_{r_x}\bm{m}(\bm{r})\cdot\te{\chi}\cdot\bm{\hat{y}}+m_z(\bm{r})(\partial_{r_y}w_x-\partial_{r_x}w_y)\bigg) \nonumber \\
    \times &\int\dfrac{d^3k}{(2\pi)^3}\bigg(\partial_T f^0[\varepsilon_l]\mathscr{Q}(\bm{k})\bigg)\bigg]
\end{align}

\noindent
For periodic magnetic textures this vanishes such that terms deriving from Eq. \eqref{aE2app} will scale at orders higher than $(\lambda/\mathcal{E}_F)(a/L_s)$.  We note terms in the integrand over real space transform trivially under rotations of the system, but flip sign under vertical mirrors. To proceed we look at terms of order $(a/L_s)^2$.

\subsubsection{Second Order in $a/L_s$}

Unlike at first order in $a/L_s$, at second order in $a/L_s$ there are non-vanishing contributions to the transverse Nernst conductivity that are order $(\lambda/\mathcal{E}_F)^0=1$ such that we may set $\lambda=0$ in Eq. \eqref{aE2} and look for the largest terms.  In the absence of SOC the equation of motion for the wave packets are

\begin{align}
    \dot{r}^{\phantom{\dagger}}_\delta(\bm{\xi})\bigg|_{\lambda=0}&=\dfrac{\partial_{k^{\phantom{\dagger}}_\delta}\varepsilon(k)}{\hbar}\nonumber \\
    \dot{k}^{\phantom{\dagger}}_\delta(\bm{\xi})\bigg|_{\lambda=0}&=\sum_{\gamma}[\te{\Omega}_{rr}]_{\delta\gamma}\bigg|_{\lambda=0} \dfrac{\partial_{k^{\phantom{\dagger}}_\gamma} \varepsilon(k)}{\hbar} \nonumber \\
    \dot{r}_z(\bm{\xi})\bigg|_{\lambda=0}&=\dfrac{\partial_{k_z}\varepsilon(k)}{\hbar},\,\,\, \dot{k}_z=0
\end{align}

\noindent
where again for simplicity in the above and what follows gradients and vectors are restricted to the $xy$-plane.  Substitution into Eq. \eqref{aE2} gives

\begin{align}
    \alpha_H^{T}&=\dfrac{e\tau^2}{2}\sum_{l=\pm}\int \dfrac{d^6\xi}{V(2\pi)^3}\partial_Tf^0[\varepsilon_l(k)]\dfrac{\partial_{k_x}\varepsilon_l}{\hbar}\bigg[\sum_{\gamma\delta}\bigg([\te{\Omega}_{rr}]_{\delta\gamma}]\bigg|_{\lambda=0}\dfrac{\partial_{k^{\phantom{\dagger}}_\delta}\varepsilon_l}{\hbar}\bigg)\partial_{k^{\phantom{\dagger}}_\gamma}\bigg]\dfrac{\partial_{k_y}\varepsilon_l}{\hbar}-(y\leftrightarrow x) \nonumber \\
    &=-\dfrac{e\tau^2}{\hbar^3}\,\sum_{l=\pm}\bigg(\int \dfrac{d^3r}{2\pi V}\Omega^l_{r_xr_y}\bigg|_{\lambda=0}\bigg) \,\int \dfrac{d^2k}{8\pi^2}\partial_{T}f^0[\varepsilon_l]\,\mathscr{Q}(\bm{k})
\end{align}

\noindent
with

    \begin{equation}
\mathscr{Q}(\bm{k})= 1/(a^2 t^3) \bigg(\bm{v}^T\cdot(\te{M}^{-1}-{\rm Tr}(\te{M}^{-1})\mathds{1})\cdot\bm{v}\bigg)
    \end{equation}

\noindent
where $\bm{v}=(\partial_{k_x}\varepsilon,\partial_{k_y}\varepsilon$) and $\te{M}^{-1}_{k_ik_j}=\partial_{k_i}\partial_{k_j} \varepsilon_l$ with $i,j\in\{x,y\}$.  We note that 

\begin{align}
    \int \dfrac{d^2r}{2\pi V}\Omega^l_{r_xr_y}\bigg|_{\lambda=0}&=-l \int \dfrac{d^2r}{4\pi V}\bm{m}(\bm{r})\cdot(\partial_{r_x}\hat{\bm{m}}(\bm{r})\times\partial_{r_y}\hat{\bm{m}}(\bm{r})) \nonumber \\
    &=-l\, n_{\text{top}}
\end{align}

\noindent
where $n_{\text{top}}$ is the topological charge density.  Substitution into the above gives the topological thermoelectric tensor

\begin{align}
    \alpha_H^{T}&=
\dfrac{e\tau^2 t^3}{\hbar^3}\,n_{\text{top}}a^2 \sum_{l=\pm}\, l\,\int \dfrac{d^2k}{8\pi^2}\partial_{T}f^0[\varepsilon_l]\,\mathscr{Q}(\bm{k})
\end{align}

\section{The Mott Relation} \label{appD}

In the presence of an external electric field in the $xy$-plane the semiclassical equation of motion are altered.

\begin{align}
    \dot{r}_{\gamma}(\bm{\xi})&=\dfrac{1}{\mathcal{D}(\bm{\xi})}\sum_{\delta}\bigg([\te{\Omega}_{kk}]_{\gamma\delta} \bigg(\dfrac{\partial_{r^{\phantom{\dagger}}_\delta} \widetilde{\mathcal{E}}(\boldsymbol\xi)}{\hbar}+e E_\delta\bigg)+[(\mathds{1}+\te{\Omega}_{rk})]_{\gamma\delta}\dfrac{\partial_{k^{\phantom{\dagger}}_\delta}\widetilde{\mathcal{E}}(\boldsymbol\xi)}{\hbar}\bigg) \nonumber \\
    \dot{k}_\gamma(\bm{\xi})&=\dfrac{1}{\mathcal{D}(\bm{\xi})}\sum_{\delta}\bigg([\te{\Omega}_{rr}]_{\gamma\delta} \dfrac{ \partial_{k^{\phantom{\dagger}}_\delta} \widetilde{\mathcal{E}}(\boldsymbol\xi)}{\hbar}-[(\mathds{1}+\te{\Omega}_{rk}^T)]_{\gamma\delta}\bigg(\dfrac{\partial_{r^{\phantom{\dagger}}_\delta} \widetilde{\mathcal{E}}(\boldsymbol\xi)}{\hbar}+eE_\delta\bigg)\bigg)
    \\
    \dot{r}_z&=\dfrac{1}{\mathcal{D}(\bm{\xi})} \dfrac{\partial_{k_z}\tilde{\mathcal{E}}(\bm{\xi})}{\hbar},\,\,\,\dot{k}_z=0
    \label{xidot2}
\end{align}

\noindent
Whereas temperature gradient perturbations only enter Boltzmann's equation through $\bm{\nabla}_r$:

\begin{align}
    -\dfrac{f-f^0[\widetilde{\mathcal{E}}]}{\tau}=(\dot{\bm{r}}(\bm{\xi})\cdot \bm{\nabla}'_r+\dot{\bm{k}}(\bm{\xi})\cdot\bm{\nabla}_k)f(\bm{\xi})+\dot{\bm{r}}\cdot\bm{\nabla}_r T(\bm{r})\partial_T f(\bm{\xi})
    \label{boltzT}
\end{align}

\noindent
where $\bm{\nabla}'_r=\bm{\nabla}_r \hat{\bm{m}}(\bm{r})\cdot \bm{\nabla}_m$.  For electric field perturbations we may write the Boltzmann equation as

\begin{align}
    -\dfrac{f-f^0[\widetilde{\mathcal{E}}]}{\tau}&=
    (\dot{\bm{r}}_E(\bm{\xi})\cdot \bm{\nabla}'_r+\dot{\bm{k}}_E(\bm{\xi})\cdot\bm{\nabla}_k)f(\bm{\xi}) \nonumber\\
    &=(\dot{\bm{r}}(\bm{\xi})\cdot \bm{\nabla}'_r+\dot{\bm{k}}(\bm{\xi})\cdot\bm{\nabla}_k)f(\bm{\xi})+\sum_{\gamma\delta}\dfrac{1}{ \mathcal{D}}\bigg[ \bigg(\dfrac{ [\te{\Omega}_{kk}]_{\delta\gamma} } {\hbar} eE_\delta\bigg) \cdot\bm{\nabla}'_{r^{\phantom{\dagger}}_\gamma} f^0[\widetilde{\mathcal{E}}]- \bigg([\mathds{1}+\te{\Omega}_{rk}^T]_{\delta\gamma}\dfrac{eE_\delta}{\hbar}\bigg)\cdot\bm{\nabla}_{k^{\phantom{\dagger}}_\gamma} f^0[\widetilde{\mathcal{E}}]\bigg] \nonumber \\
    &=(\dot{\bm{r}}(\bm{\xi})\cdot \bm{\nabla}'_r+\dot{\bm{k}}(\bm{\xi})\cdot\bm{\nabla}_k)f(\bm{\xi})-\bm{\dot{r}} \cdot e\bm{E}\partial_{\widetilde{\mathcal{E}}}f^0[\widetilde{\mathcal{E}}]
    \label{boltzE}
\end{align}

\noindent
Comparing Eq. \eqref{boltzT} to Eq. \eqref{boltzE},
in the presence of electric field, the perturbations to the distribution function linear order in the electric field, $g^{\phi}(\bm{\xi})$ take a similar form to temperature gradient induced perturbations.  Solving the Boltzmann equation gives \cite{verma2022unified}

\begin{equation}
   g^{\phi}(\bm{\xi})=\tau e\sum_{n=0}^\infty (-1)^n(\mathds{P})^n\bigg(\dot{\bm{r}}(\bm{\xi})\cdot\bm{E}\,\partial_{\widetilde{\mathcal{E}}} f^0[\widetilde{\mathcal{E}}(\bm{\xi})]\bigg)
\end{equation}

\noindent
such that these contributions to the electric conductivity can be written as

\begin{equation}
    \sigma^{(2)}_{ij}=-e^2\tau\sum_{l=\pm}\int \dfrac{d^6\xi}{V(2\pi)^3}\mathcal{D}_l(\bm{\xi})\partial_{\widetilde{\mathcal{E}}} f^0[\widetilde{\mathcal{E}}_l(\bm{\xi})] \bigg[\dot{r}^i_l(\bm{\xi})\bigg(\sum_{n=0}^\infty (-1)^n(\mathds{P}_l)^n \bigg)\dot{r}^j_l(\bm{\xi})\bigg]
\end{equation}

\noindent
which we may compare to the contributions to the thermoelectric conductivity

\begin{equation}
    \alpha_{ij}^{(2)}=-e\tau\sum_{l=\pm}\int \dfrac{d^6\xi}{V(2\pi)^3}\mathcal{D}_l(\bm{\xi})\partial_Tf^0_l[\widetilde{\mathcal{E}}_l(\bm{\xi})] \bigg[\dot{r}^i_l(\bm{\xi})\bigg(\sum_{n=0}^\infty (-1)^n(\mathds{P}_l)^n \bigg)\dot{r}^j_l(\bm{\xi})\bigg]
\end{equation}

\noindent
For simplicity we define

\begin{equation}
G_l(\bm{\xi})=-e\tau\dfrac{1}{V(2\pi)^3}\mathcal{D}_l(\bm{\xi})\bigg[\dot{r}^i_l(\bm{\xi})\bigg(\sum_{n=0}^\infty (-1)^n(\mathds{P}_l)^n \bigg)\dot{r}^j_l(\bm{\xi})\bigg]
\end{equation}

\noindent
such that we may write the conductivities as

\begin{align}
    \alpha_{ij}^{(2)}&=\sum_{l=\pm}\int d^6\xi \,\, \partial_Tf^0[\widetilde{\mathcal{E}}_l(\bm{\xi})] G_l(\bm{\xi})=-\sum_{l=\pm}\int d^6\xi \,\,  \bigg(\dfrac{\widetilde{\mathcal{E}}(\bm{\xi})-\mu}{k_BT}\bigg)\partial_{\widetilde{\mathcal{E}}_l} f^0[\widetilde{\mathcal{E}}_l(\bm{\xi})] G_l(\bm{\xi}) \nonumber \\
     \sigma_{ij}^{(2)}&=e\sum_{l=\pm}\int d^6\xi \,\,  \partial_{\widetilde{\mathcal{E}}_l} f^0[\widetilde{\mathcal{E}}_l(\bm{\xi})] G_l(\bm{\xi})
\end{align}

\noindent
We may define

\begin{equation}
    \tilde{G}_l(x)=\int d^6\xi \,\, \delta(\widetilde{\mathcal{E}}_l(\bm{\xi})-x) G_l(\bm{\xi})
\end{equation}

\noindent
such that we may write the conductivities as

\begin{align}
    \alpha_{ij}^{(2)}&=\sum_{l=\pm}\int dx \,\, \partial_{x} f^0[x] \tilde{G}_l(x) \bigg(\dfrac{x-\mu}{k_BT}\bigg) \nonumber \\
     \sigma_{ij}^{(2)}&=-e\sum_{l=\pm}\int dx \,\,  \partial_{x} f^0[x] \tilde{G}_l(x)
\end{align}

\noindent
For low temperatures we can now do a Sommerfeld expansion of both $\sigma^{E}_{ij}$ and $\alpha^{E}_{ij}$

\begin{align}
    \alpha_{ij}^{(2)}&\approx-\dfrac{\pi^2}{3}k_B^2T\sum_{l=\pm} \dfrac{\partial \tilde{G}_l(x)}{\partial x}\bigg|_{x=\mu} \nonumber \\
    \sigma_{ij}^{(2)}&\approx e\sum_{l=\pm}\tilde{G}_l(\mu)
\end{align}

\noindent
which implies the Mott relation

\begin{equation}
\alpha_{ij}^{(2)}\approx -\dfrac{\pi^2}{3}\dfrac{k_B^2T}{e}\dfrac{\partial \sigma_{ij}^{(2)}}{\partial \mu}
\label{MottRel2}
\end{equation}

\section{Chemical Potential Dependence of Anomalous Hall Conductivity}\label{appE}

The anomalous Hall conductivity is proportional to the integral of the momentum space Berry curvature over the occupied Bloch states.  Its leading order contribution is

\begin{align}
    \sigma^{A}_{H}&=-\dfrac{e^2}{2\hbar}\sum_l \dfrac{l}{J^2}\int \dfrac{d^3k}{(2\pi)^3}f^0[\varepsilon_l]\bigg(\partial_{k_x}\bm{d}_l(\bm{k})\times \partial_{k_y}\bm{d}_l(\bm{k})\bigg)\bigg|_{\bm{w}=0}\cdot\hat{\bm{z}}
\end{align}

\noindent
For simplicity we suppress the band index in what follows.  To see the relationship to $\mathscr{Q}(\bm{k})$ we first perform an integration be parts

\begin{align}
\int\dfrac{d^3k}{(2\pi)^3}f^0[\varepsilon]\bigg(\partial_{k_x}\bm{d}(\bm{k})\times \partial_{k_y}\bm{d}(\bm{k})\bigg)\bigg|_{\bm{w}=0}\cdot\hat{\bm{z}}=-\dfrac{t}{2}\int\dfrac{d^3k}{(2\pi)^3} \sum_{i,j,n,m=x,y}\varepsilon_{ij}\varepsilon_{nm} d_n (\partial_{k_i} d_m) v_j \partial_\varepsilon f^0[\varepsilon]
\label{EsigAH}
\end{align}

\noindent
where $\varepsilon_{ij}$ is the Levi-Civita symbol which in two dimensions has a single independent coefficient.  The product allows four nonzero terms that can be group as follows

\begin{align}
\sum_{i,j,n,m} \varepsilon_{ij}\varepsilon_{nm} d_n (\partial_{k_i} d_m) v_j\bigg|_{\bm{w}=0}&=-\dfrac{1}{a^2}\text{Det}(\te{\chi})\bigg(-v_x(\partial_{k_y}v_x) v_y-v_y(\partial_{k_x}v_y)v_x+\partial_{k_x}v_xv_y^2+\partial_{k_y}v_yv_x^2\bigg) \nonumber \\
&= \dfrac{1}{a^2}\text{Det}(\te{\chi})\mathscr{Q}(\bm{k})
\end{align}

\noindent
Substitution into Eq. \eqref{EsigAH} gives Eq. \eqref{sigAH}.

To understand why $\mathscr{Q}(\bm{k})$ appears in both the anomalous and topological contributions, we note that the totally anti-symmetric part of any rank two tensor must be invariant under rotations about the $\hat{\bm{z}}$-axis $\te{R}^z(\theta)$ and must change sign under vertical mirror planes, $\te{\mathcal{M}}^x$ and $\te{\mathcal{M}}^y$.  For any rank two space tensor $\te{T}$, under rotations $T_{ij}\rightarrow T_{ij}'=\sum_{kl}R^z_{ik}(\theta)R^z_{jl}(\theta)T_{kl}$.  The totally anti-symmetric component must be left invariant as $T'_{xy}-T'_{yx}=T_{xy}-T_{yx}$.  We also note that under vertical mirror planes, $\mathcal{M}^x$ and $\mathcal{M}^y$, the totally anti-symmetric component must change sign: $T_{ij}\rightarrow T_{ij}'=\sum_{kl}\mathcal{M}^x_{ik}\mathcal{M}^x_{jl}T_{kl}=\sum_{kl}\mathcal{M}^y_{ik}\mathcal{M}^y_{jl}T_{kl}=-T_{ij}$.  In $\alpha_H^T$, these mirror operations flip the sign of $n_{\text{sk}}$.  In $\alpha_H^A$, it is $\bar{m}_z$ which changes sign under such a transformation, while $\text{Det}(\te{\chi})$ is left invariant.

\section{Calculation of Thermoelectric Transport and the Sondheimer Cancellation}\label{appF}

For quadratic dispersion, the conductivities $\sigma_L$, $\sigma^A_H$, and $\sigma_H^T$ derive from contributions from each band that scale linearly with the electronic density or chemical potential.  The anomalous and topological contributions scale linearly with the Berry curvature such that their values are oppositely signed in each band such that the two terms contributing to $\mathscr{A}(\mu)$ have opposite sign.  This leads to a complete cancellation in $\mathscr{A}(\mu)$ when electronic states of both bands are filled. 
 For the Nernst signal $\mathscr{N}(\mu)= -4\pi^3 t J/3\mu^2 \, \Theta(\mu-(\varepsilon_0+J))$ is nonzero only when electron states in both bands are occupied (see Fig. \ref{nernstalphaPlot}(b)).  When the Fermi surface contains a single electron pocket, the longitudinal, topological Hall, and anomalous Hall electric conductivity scale linearly with $\mu$ such that $\mathscr{N}(\mu)$ and the Nernst effect vanish.  However, when the Fermi surface contains two electron pockets, the anomalous and topological Hall conductivities are independent of $\mu$, as the curvatures are oppositely signed in the two bands.  In contrast, the longitudinal conductivity scales linearly with $\mu$, leading to $\mathscr{N}(\mu)\propto 1/\mu^2$ for $\mu>\varepsilon_0+J$ and is non-divergent as $\mu$ approaches the band edge.  In comparison,  $S_L$ for parabolic bands is approximately divergent as $\mu$ approaches the band edge (see Appendix \ref{appF}).  Of course exactly at the band edge the Mott relation is invalid for any temperature and it is better to calculate $S_L=\sigma_L^{-1}\alpha_L$ using the exact expression for the longitudinal component of the thermoelectric conductivity (Eq. \eqref{alphaL}) and the electric conductivity.  

In canonical thermoelectric transport in the presence of external magnetic fields a Sondheimer cancellation can occurs whereby the Nernst effect is suppresses due to a constant Hall angle independent of a system's chemical potential.  This occurs when the chemical potential dependence of the longitudinal conductivity and Hall conductivity are equivalent.  For weakly spin-orbit coupled spin-split bands the Nernst effect in the presence of an external magnetic field is determined by equation \eqref{NernstMu} with $l\rightarrow 1$ in the numerator and $n_{\text{top}}$ replaced by $B e/h$.  For quadratic bands in three dimensions this leads to $\sigma^B_{H},\sigma^B_L\propto  (\mu+J)^{3/2}+(\mu-J)^{3/2}$ such that $N\propto\partial_\mu \Theta_H=0$.  Whereas in the presence of nontrivial topological charge $l=\pm 1$ as the sign of $\Omega_{r_xr_y}$ is opposite for each spin split pair of bands leading to $\sigma^{\Omega_{rr}}_{L}\propto (\mu+J)^{3/2}+(\mu-J)^{3/2}$ and $\sigma^{\Omega_{rr}}_{H}\propto (\mu+J)^{3/2}-(\mu-J)^{3/2}$ such that generically $N\propto\partial_\mu \Theta_H\neq 0$.  This allows for Nernst effects to occur even in the absence of a rapidly changing energy dependent scattering or relaxation time.

\begin{centering}
\begin{figure}[ht]
\includegraphics[width=0.7\textwidth]{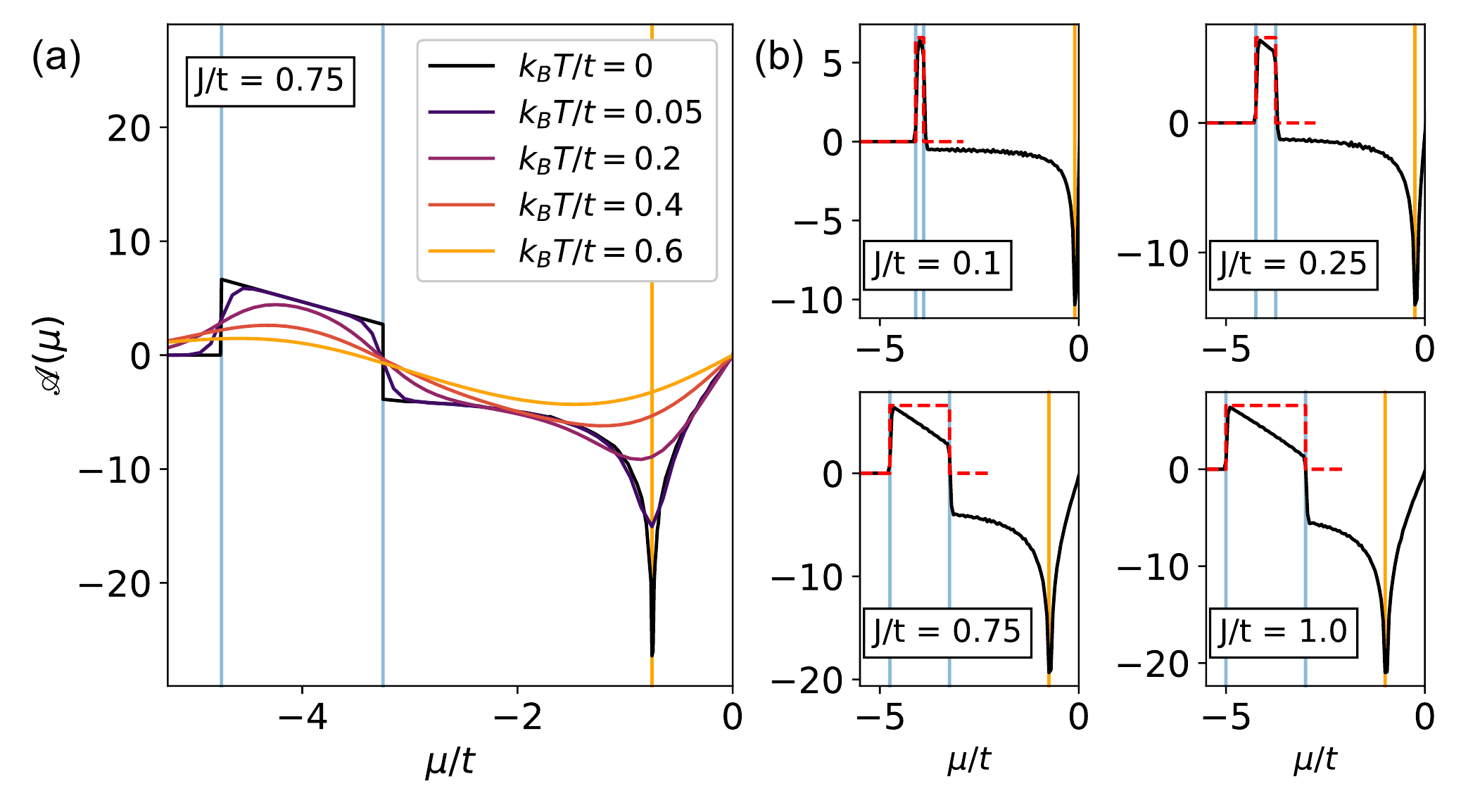}
\caption{{\bf $\mathscr{A}(\mu)$ for varying parameters.} The left panel shows the effect of raising the temperature. The finite temperature values are calculated using the temperature dependent Eq. \eqref{trueAlpha}. Non-physical jumps are smoothed out. The zero-temperature approximation is appropriate in metals and heavily doped semimetals. The right panels show the effect of varying $J/t$ while holding $T=0$. As in Fig. \ref{nernstalphaPlot}, the Van Hove singularities (gold) and band edges (blue) are shown by vertical lines. The dashed red line shows the approximate solution for quadratic bands. As $J/t$ is increased, the qualitative behavior remains the same, as long as no gap is opened between the bands.
}
\label{Alphaparams}
\end{figure}
\end{centering}

We consider nearest neighbor interactions on the 2D square lattice with spin splitting $J$. Fig. \ref{Alphaparams}a shows the temperature dependence of $\mathscr{A}(\mu)$.
The non-analytic jumps in $\mathscr{A}(\mu)$ are the result of the Mott relation, in which the $\mu$ derivative is taken at $T=0$. The full temperature-dependent expression from Eq. \eqref{trueAlpha} is plotted in comparison to the Mott result. The Mott result remains a good approximation since $k_B T \ll t$.

The energy gap between bands is proportional to the magnetic exchange coupling $J$. Fig. \ref{Alphaparams}b shows $\mathscr{A}(\mu)$ within a model of nearest neighbor interaction on the square lattice $(\varepsilon(\bm{k})=2t(\cos(k_x)+\cos(k_y)))$ for various $J/t$.  The behavior of $\mathscr{A}(\mu)$ for each choice of $J/t$ remains qualitatively the same. For small density the quadratic band approximation overestimates $\mathscr{A}(\mu)$ with increasing $\mu$ and $J/t$ such that $\mathscr{A}(\mu)$ steadily shifts away from the approximation, but still remains relatively constant in $\mu$. Additionally, the change in $\mathscr{A}(\mu)$ for $\mu$ below and above the upper band edge remains constant over this range of $J/t$. This qualitative behavior derives from the particular distribution of Berry curvature in the Brillouin zone near the $\Gamma$ point. For $\mu$ near the upper band edge, the ratio of the Berry curvature in each band, $|\Omega_{k_xk_y}^-|/|\Omega_{k_xk_y}^+|$, can be shown to decrease with $J/t$.

In two dimensions and using the Mott relation, the analytic expressions for the longitudinal thermoelectric conductivity for quadratic bands is given by

\begin{align}
\alpha_L&\approx\dfrac{\pi^3}{6}\dfrac{k_Be}{\hbar}\bigg(\dfrac{k_BT}{
(\hbar/\tau)}\bigg)
\begin{cases}
1 & \varepsilon_0-J<\mu< \varepsilon_0+J \\
2 & \mu>\varepsilon_0+J
\end{cases}
  \label{alphaT}
\end{align}

\noindent
Using the Mott relation and in the limit $\sigma_L=(\sigma_{xx}+\sigma_{yy})/2\gg (\sigma_{xx}-\sigma_{yy})/2,\sigma_{xy},\sigma_{yx}$ the analytic expression for leading order in $T$ contribution to the Seebeck effect or thermopower is

\begin{align}
    S_L&\approx -\dfrac{\pi^2}{3}\dfrac{k_B^2T}{e}\dfrac{\partial}{\partial \mu}\bigg(\ln(\sigma_L)\bigg) =-\dfrac{\pi^2}{3}\dfrac{k_B^2T}{e}\bigg(\dfrac{\Theta[(\mu-(\varepsilon_0-J)]}{\mu-(\varepsilon_0-J)}+\dfrac{\Theta[(\mu-(\varepsilon_0+J)]}{\mu-(\varepsilon_0+J)}\bigg)
\end{align}

\end{document}